\documentclass{article}
\pdfoutput=1
\usepackage{graphicx}  
\usepackage{amsmath}   
\usepackage[compress]{cite}
\usepackage{amssymb}   
\usepackage{bm} 
\usepackage{dcolumn}
\usepackage{color}
\usepackage{mathrsfs}
\usepackage{amsfonts}
\usepackage{varioref}
\usepackage{mathrsfs}
\usepackage{graphicx}
\usepackage{latexsym}
\usepackage{amsmath}
\usepackage{amssymb}
\usepackage{textcomp}
\usepackage{amsbsy}
\usepackage{graphics}
\usepackage{epstopdf}
\usepackage{color}
\usepackage[caption=false]{subfig}
\RequirePackage[colorlinks,citecolor=blue,urlcolor=magenta,linkcolor=blue]{hyperref}
\input epsf

\labelformat{section}{Section #1} 
\labelformat{subsection}{Section #1} 
\labelformat{subsubsection}{Section #1}
\labelformat{subsubsubsection}{Section #1}
\labelformat{equation}{Eq.~(#1)} 
\labelformat{figure}{Fig.~#1} 
\labelformat{subfigure}{Fig.~\thefigure#1} 
\labelformat{table}{Tab.~#1} 
\labelformat{appendix}{Appendix #1}
\title{Unifying an asymmetric bounce to the dark energy in Chern-Simons F(R) gravity}

\author{Sergei~D.~Odintsov$^{1,2}$\footnote{odintsov@ieec.uab.es}~,
Tanmoy~Paul$^{3,4}$\footnote{pul.tnmy9@gmail.com~(Corresponding author)}~,
Indrani~Banerjee$^{5}$\footnote{banerjeein@nitrkl.ac.in}~,
Ratbay~Myrzakulov$^{6,7}$\footnote{rmyrzakulov@gmail.com}~\\
and~Soumitra~SenGupta$^{8}$\footnote{tpssg@iacs.res.in} \\
\small{$^{1}$ ICREA, Passeig Luis Companys, 23, 08010 Barcelona, Spain}\\
\small{$^{2}$ Institute of Space Sciences (IEEC-CSIC) C. Can Magrans s/n,08193 Barcelona, Spain}\\
\small{$^{3}$ Department of Physics, Chandernagore College, Hooghly - 712 136, India}\\
\small{$^{4}$ International Laboratory for Theoretical Cosmology, TUSUR, 634050 Tomsk, Russia}\\
\small{$^{5}$Department of Physics and Astronomy, National Institute of Technology, Rourkela-769008, India }\\
\small{$^{6}$ Ratbay Myrzakulov Eurasian International Centre for Theoretical Physics, Nur-Sultan 010009, Kazakhstan}\\
\small{$^{7}$ Eurasian  National University, Nur-Sultan 010008, Kazakhstan}\\
\small{$^{8}$ School of Physical Sciences, Indian Association for the Cultivation of Science, Kolkata-700032, India}}

\date{ }  
\begin{document}
  
\maketitle 
\begin{abstract}
We propose a cosmological scenario in which the universe undergoes through a non-singular bounce, and after the bounce, it decelerates 
having a matter-like dominated evolution during some regime of the deceleration era, and finally at the present epoch it evolves through 
an accelerating stage. Our aim is to study such evolution in the context of Chern-Simons corrected F(R) gravity theory and confront the model 
with various observational data. Using the reconstruction technique, and in addition by employing suitable boundary conditions, we determine 
the form of F(R) for the entire possible range of the cosmic time. The form of F(R) seems to unify a non-singular bounce with a dark energy epoch, in 
particular, from a non-singular bounce to a deceleration epoch and from a deceleration epoch to a late time acceleration era. It is important to mention 
that the bouncing scenario in the present context is an asymmetric bounce, in particular, the Hubble radius monotonically increases and asymptotically 
diverges at the late contracting era, while it seems to decrease with time at the present epoch. The decreasing behaviour of the Hubble radius ensures a 
late time acceleration era of the universe. Moreover, due to the aforesaid evolution of the Hubble radius, the primordial perturbation modes 
generate at the deep contracting era far away from the bounce when all the perturbation modes lie within the horizon. Correspondingly 
we calculate the scalar and tensor power spectra, and accordingly, we evaluate the primordial observable quantities like the spectral index of the 
scalar curvature perturbation, the tensor-to-scalar ratio, and as a result, they are found to be in agreement with the latest Planck 2018 constraints. 
In this regard, the Chern-Simons term seems to have considerable effects on the tensor perturbation evolution, 
however keeping intact the scalar part of the perturbation with that of in the case of a vacuum F(R) model, and as a result, the Chern-Simons term 
proves to play an important role in making the observable quantities consistent with the Planck results. Furthermore the theoretical expectation 
of the effective equation of state parameter of the dark energy epoch is confronted with the Planck+SNe+BAO data.

\end{abstract}
\begin{itemize}
 \item \textbf{Keywords}: {Unification of cosmological epochs, Bouncing cosmology, Dark energy epoch, Cosmological perturbation, 
 Higher curvature gravity theory, Planck data.}
\end{itemize}

\section{Introduction}
\label{Intro}
Cosmology today is largely data-driven which opens up the opportunity to construct a  consistent history of the universe that explains the observations. 
In this light, it is intriguing that the current observations and experiments delimits the evolution of the universe 
in terms of the well established laws of physics, e.g. general relativity or some modified theories of gravity, 
the Standard Model of particle physic, fluid dynamics etc. However, direct experimental results probing physics above the 
TeV scale ceases to exist which also turns out to be a major impediment towards an unambiguous understanding of the physics of 
the very early universe. This has given birth to diversified early universe scenarios, e.g. the \emph{inflationary scenario} 
\cite{guth,Linde:2005ht,Langlois:2004de,Riotto:2002yw,barrow1,barrow2,Baumann:2009ds}, 
the \emph{bouncing universe} \cite{Brandenberger:2012zb,Brandenberger:2016vhg,Battefeld:2014uga,Novello:2008ra,Cai:2014bea,Cai:2016thi,
deHaro:2015wda,Lehners:2011kr,Lehners:2008vx,
Cai:2016hea,Colin:2017dwv,Cattoen:2005dx,Li:2014era,Brizuela:2009nk,Cai:2013kja,Quintin:2014oea,Cai:2013vm,pinto,
Koehn:2015vvy,Nojiri:2016ygo,Odintsov:2020zct,Koehn:2013upa,Battarra:2014kga,Martin:2001ue,Khoury:2001wf,
Hackworth:2004xb,Johnson:2011aa,Peter:2002cn,Gasperini:2003pb,Creminelli:2004jg,Lehners:2015mra,
Lehners:2013cka,Cai:2014xxa,Cai:2007qw,Barrow:2004ad,Haro:2015zda,Das:2017jrl,
Cai:2008qw,Finelli:2001sr,Cai:2011ci,Haro:2015zta,Cai:2011zx,Brandenberger:2009yt,deHaro:2014kxa,Odintsov:2014gea,
Qiu:2010ch,Bamba:2012ka,deHaro:2012xj,Nojiri:2019lqw,Elizalde:2019tee,Elizalde:2020zcb,WilsonEwing:2012pu}, 
the \emph{emergent universe scenario} \cite{Ellis:2003qz,Paul:2020bje,Li:2019laq} etc, all of which can consistently explain the nearly 
scale-invariant power spectrum or the low tensor to scalar ratio observed by the Planck satellite \cite{Akrami:2018odb}.  

In the present work, we take the route of bouncing scenario which comprises of an era of contraction 
followed by an era of expansion of the scale factor, both the epochs being connected by a non-singular bounce. Such a scenario is 
interesting, particularly in the absence of a successful quantum theory of gravity, as it evades the Big Bang singularity which is indeed an 
unavoidable feature of inflationary scenario in the realm of 
classical gravity when extrapolated backwards in time. 

The bouncing scenario has been studied extensively in the context of scalar tensor models and other modified gravity theories 
\cite{Brandenberger:2012zb,Brandenberger:2016vhg,Battefeld:2014uga,Novello:2008ra,Cai:2014bea,Cai:2016thi,
deHaro:2015wda,Lehners:2011kr,Lehners:2008vx,
Cai:2016hea,Colin:2017dwv,Cattoen:2005dx,Li:2014era,Brizuela:2009nk,Cai:2013kja,Quintin:2014oea,Cai:2013vm,pinto,
Koehn:2015vvy,Nojiri:2016ygo,Odintsov:2020zct,Koehn:2013upa,Battarra:2014kga,Martin:2001ue,Khoury:2001wf,
Hackworth:2004xb,Johnson:2011aa,Peter:2002cn,Gasperini:2003pb,Creminelli:2004jg,Lehners:2015mra,
Lehners:2013cka,Cai:2014xxa,Cai:2007qw,Barrow:2004ad,Haro:2015zda,Das:2017jrl,
Cai:2008qw,Finelli:2001sr,Cai:2011ci,Haro:2015zta,Cai:2011zx,Brandenberger:2009yt,deHaro:2014kxa,Odintsov:2014gea,
Qiu:2010ch,Bamba:2012ka,deHaro:2012xj,Nojiri:2019lqw,Elizalde:2019tee,Elizalde:2020zcb,WilsonEwing:2012pu}, 
some of which are often inspired 
from string theory. Among various bounce models proposed so far, the matter bounce scenario (MBS) 
\cite{deHaro:2015wda,Cai:2008qw,Finelli:2001sr,Quintin:2014oea,Cai:2011ci,
Haro:2015zta,Cai:2011zx,Cai:2013kja,Brandenberger:2009yt,deHaro:2014kxa,Odintsov:2014gea,
Qiu:2010ch,Bamba:2012ka,deHaro:2012xj,Nojiri:2019lqw,Elizalde:2019tee,Elizalde:2020zcb,
WilsonEwing:2012pu} earned a lot of attention as it 
generates an almost scale invariant primordial power spectrum and also leads to a matter dominated epoch during the late expanding phase. 
Moreover in the matter bounce theory, the Universe evolved from an epoch at large negative time in the contracting era 
where the primordial spacetime perturbations are generated deeply inside the Hubble radius, and is thus able to solve the horizon problem. 
Despite these successes, the matter bounce scenario hinges with some serious problems, in particular, an exact MBS characterized by a single scalar field 
leads to an \emph{exactly} scale invariant power spectrum or equivalently 
a vanishing running of the spectral index and a tensor-to-scalar ratio of order unity,  all of which are 
inconsistent with the latest Planck results. Such problematic issues was also confirmed from a different point of view, in particular in the context of 
F(R) gravity in \cite{Odintsov:2014gea,Nojiri:2019lqw} (for early and late time cosmology in F(R) gravity, see \cite{Nojiri:2017ncd,Nojiri:2010wj,
Elizalde:2018rmz}). In this regard, we would like to mention that a scalar-tensor model can be equivalently 
mapped to F(R) gravity by a suitable conformal transformation 
of the spacetime metric and thus the inconsistencies of MBS with the Planck observations in both the scalar-tensor and F(R) model 
are well justified. Furthermore in both the matter bounce or quasi-matter bounce scenarios, the comoving Hubble radius 
(defined by $r_h = 1/\left(aH\right)$) monotonically increases with time and diverges to 
infinity at the distant future, which in turn leads to a deceleration stage of the universe at the late expanding phase and thus 
fails to explain the dark energy epoch of the universe as expected from SNe-Ia+BAO+H(z)+CMB observations 
\cite{Perlmutter:1996ds,Perlmutter:1998np,Riess:1998cb}. 

Some of the above mentioned problems seem to be cured when the background bounce scenario is considered to be a 
quasi-matter bounce scenario where the FRW scale factor behaves as quasi-matter dominated epoch during the asymptotic time. 
In particular, the scalar-tensor of F(R) quasi-matter bounce model yields a \emph{nearly} 
scale invariant power spectrum (in accordance with the Planck results), although the amplitudes of tensor and scalar perturbations 
continue to be comparable, due to which the tensor-to-scalar ratio remains to be order of unity \cite{deHaro:2015wda}. In order to resolve the 
issue of the high tensor-to-scalar ratio, some suitable higher curvature gravity theories come into the picture, with success in many of the 
cases \cite{Nojiri:2019lqw,Elizalde:2020zcb} (for a general review on modified gravity, see \cite{Nojiri:2017ncd,Nojiri:2010wj,Capozziello:2011et,
Capozziello:2010zz}). 
However in most modified gravity theories describing the bouncing cosmology, the comoving Hubble radius increases with 
cosmic time and leads to a deceleration era of the universe at distant future, and thus the problem of describing a dark energy epoch 
still persists in such bouncing models.

Motivated by this problem, in the present work, we aim to study bouncing cosmology which is also compatible with the dark energy epoch of our 
current universe. For this purpose, we consider the Chern-Simons (CS) corrected F(R) gravity theory, where the presence of the CS coupling induces a 
parity violating term in the gravitational action. The gravitational Chern-Simons term arises in the low-energy effective action of several string inspired 
models \cite{Green:1984sg,Antoniadis:1992sa} and studying its role in explaining 
the primordial power spectrum may provide an indirect testbed for string theory. 
The parity violating Chern-Simons gravity distinguishes the evolution of the two polarization modes of primordial gravitational 
waves \cite{Hwang:2005hb,Choi:1999zy}, which leads to the generation of 
chiral gravitational waves leaving non-trivial imprints in the Cosmic 
Microwave Background Radiation (CMBR). Such signatures if detected in the future generation of experiments may signal the presence of the string 
inspired Chern-Simons gravity in the early universe. This has motivated several works in this direction, see for example 
\cite{Satoh:2007gn,Satoh:2008ck,Jackiw:2003pm,Lue:1998mq,Choi:1999zy,Haghani:2017yjk,Nishizawa:2018srh,Inomata:2018rin,Kamionkowski:1997av}. 
The astrophysical implications of the gravitational Chern-Simons (GCS) term has also been explored e.g. \cite{Wagle:2018tyk}.
A further and important motivation to include the CS term in the present context of F(R) gravity 
stems from the fact that, as mentioned earlier, vacuum F(R) bounce models generally cannot reproduce the observed tensor-to-scalar ratio in 
respect to the Planck data. However the addition of the Chern-Simons term in the F(R) gravity will possibly 
resolve this issue as the CS term does not affect the evolution of the spatially flat FRW background or the scalar perturbations, 
but plays a pivotal role in the evolution of the tensor perturbations. Moreover some of our authors explored the 
importance of the Chern-Simons F(R) model in producing a viable $inflationary$ scenario compatible with the Planck results \cite{Odintsov:2019mlf}. 
This further motivates us to explore the relevance of such a 
model in inducing a $bouncing$ universe and subsequently unifying it with the dark energy epoch. Based on these arguments, in the present paper, 
we try to provide a cosmological model which unifies certain cosmological era of the universe-- from a non-singular bounce to a matter dominated 
era and from the matter dominated to the dark energy epoch. Here we would like to mention that the unification of bounce with dark energy era 
has been studied earlier in \cite{Odintsov:2020zct,Odintsov:2016tar,Cai:2016hea,Colin:2017dwv}, however in a different context; for instance, in 
\cite{Odintsov:2020zct}, 
the Hubble radius asymptotically goes to zero at 
both sides of the bounce and thus the primordial perturbation modes generate near the bounce, unlike to 
the present work where the perturbation modes generate at the distant past far away from the bounce.

The paper is organized as follows: in \ref{sec-model}, 
we will briefly describe the essential features of Chern-Simons F(R) gravity theory. 
Having set the stage, \ref{sec-background} describes the background evolution and the constraints on various model parameters, 
while \ref{sec-perturbation} is reserved for studying the perturbation evolution and their confrontation with the Planck results. 
The paper will end with some conclusions along with some scope for future work.

\section{Essential features of Chern-Simons F(R) gravity}\label{sec-model}

Let us briefly recall some basic features of Chern-Simons corrected $F(R)$ gravity, which are necessary for our presentation 
\cite{Hwang:2005hb,Odintsov:2019mlf}. 
The gravitational action of $F(R)$ gravity generalized by Chern-Simons coupling is given by,
\begin{eqnarray}
 S = S = \int d^4x\sqrt{-g}\frac{1}{2\kappa^2}\left[F(R) + \frac{1}{8}\nu(R)~\tilde{R}^{\mu\nu\alpha\beta}R_{\mu\nu\alpha\beta}\right]
 \label{action}
\end{eqnarray}
where $\nu(R)$ is known as Chern-Simons coupling function, 
$\tilde{R}^{\mu\nu\alpha\beta} = \epsilon^{\gamma\delta\mu\nu}R_{\gamma\delta}^{~~\alpha\beta}$, 
$\kappa^2$ stands for $\kappa^2 = 8\pi G = \frac{1}{M_\mathrm{Pl}^2}$ and also $M_\mathrm{Pl}$ 
is the reduced Planck mass. By using the metric formalism, we vary the
action with respect to the metric tensor $g_{\mu\nu}$, and the gravitational equations read,
\begin{eqnarray}
 F'(R)R_{\mu\nu} - \frac{1}{2}F(R)g_{\mu\nu} - \nabla_{\mu}\nabla_{\nu}F'(R) + g_{\mu\nu}\Box F'(R) = T^{(c)}_{\mu\nu}~~,
 \label{basic2}
\end{eqnarray}
with
\begin{eqnarray}
 T^{(c)}_{\mu\nu}&=&
 \frac{2}{\sqrt{-g}}\frac{\delta}{\delta g^{\mu\nu}}\left\{\frac{1}{8}\sqrt{-g}~\nu(R)\tilde{R}^{\mu\nu\alpha\beta}R_{\mu\nu\alpha\beta}\right\}\nonumber\\
 &=&\nu'(R)\tilde{R}^{\mu\nu\alpha\beta}R_{\mu\nu\alpha\beta}\left(\frac{\delta R}{\delta g^{\mu\nu}}\right) 
 + \epsilon_{\mu}^{~cde}\left[\nu_{,e;f}R^{f}_{~\nu cd} - 2\nu_{,e}R_{\nu c;d}\right]
 \label{EM tensor-CS}
\end{eqnarray}
is the energy-momentum tensor contributed from the Chern-Simons term \cite{Hwang:2005hb}. Moreover 
$R_{\mu\nu}$ is the Ricci tensor constructed from $g_{\mu\nu}$ and $\nu'(R) = \frac{d\nu}{dR}$. 
Since the present article is devoted to cosmological context,
the background metric of the Universe will be assumed to be a flat Friedmann-Robertson-Walker (FRW) metric,
\begin{eqnarray}
 ds^2 = -dt^2 + a^2(t)\big[dx^2 + dy^2 + dz^2\big]
 \label{basic3}
\end{eqnarray}
with $a(t)$ being the scale factor of the Universe. For such metric, the non-zero components of the Riemann tensor are given by 
$R_{0i}^{~~0j}$ and $R_{ij}^{~~kl}$ (with $i,j,k,l$ denote the spatial indices), in particular, the FRW metric \ref{basic3} leads to the Ricci 
scalar and the non-zero components of Ricci tensor, Riemann tensor as,
\begin{eqnarray}
 R&=&6\left(\frac{\ddot{a}}{a} + \frac{\dot{a}^2}{a^2}\right)~~~~,~~~~R_{00} = -3\ddot{a}/a~~~~,~~~~
 R_{ij} = \left(a\ddot{a} + 2\dot{a}^2\right)\delta_{ij}\nonumber\\
 R_{0i}^{~~0j}&=&-\left(\frac{\ddot{a}}{a}\right)\delta^{j}_{i}~~~~~,~~~~~
 R_{ij}^{~~kl} = \frac{\dot{a}^2}{a^2}\left(\delta_{i}^{l}\delta_{j}^{k} - \delta_{i}^{k}\delta_{j}^{l}\right)
 \nonumber
\end{eqnarray}
respectively, where an overdot represents $\frac{d}{dt}$. 
In effect, the energy-momentum tensor $T^{(c)}_{\mu\nu}$ identically vanishes in the background of FRW spacetime, i.e we may argue that 
the Chern-Simons term does not affect the background Friedmann equations, as also stressed in \cite{Hwang:2005hb}. However as we will see later that the 
Chern-Simons term indeed affects the perturbation evolution over the FRW spacetime, particularly the tenor type perturbation. Hence 
the temporal and spatial components of \ref{basic2} become,
\begin{eqnarray}
0&=&-\frac{F(R)}{2} + 3\big(H^2 + \dot{H}\big)F'(R) - 18\big(4H^2\dot{H} + H\ddot{H}\big)F''(R)\nonumber\\
0&=&\frac{F(R)}{2} - \big(3H^2 + \dot{H}\big)F'(R) + 6\big(8H^2\dot{H} + 4\dot{H}^2 + 6H\ddot{H}
+ \dddot{H}\big)F''(R) + 36\big(4H\dot{H} + \ddot{H}\big)^2F'''(R)~~,
\label{basic4}
\end{eqnarray}
where $H = \dot{a}/a$ denotes the Hubble parameter of the 
Universe. Comparing \ref{basic4} with the usual Friedmann equations, it is easy to 
reveal that the F(R) gravity provides an effective energy-momentum tensor with the following forms of 
effective energy density ($\rho_\mathrm{eff}$) and effective pressure ($p_\mathrm{eff}$),
\begin{eqnarray}
 \rho_\mathrm{eff} = \frac{1}{\kappa^2}\bigg[-\frac{f(R)}{2} + 3\big(H^2 + \dot{H}\big)f'(R) - 18\big(4H^2\dot{H} + H\ddot{H}\big)f''(R)\bigg]
 \label{ed}
\end{eqnarray}
\begin{eqnarray}
 p_\mathrm{eff} = \frac{1}{\kappa^2}\bigg[\frac{f(R)}{2} - \big(3H^2 + \dot{H}\big)f'(R) + 6\big(8H^2\dot{H} + 4\dot{H}^2 + 6H\ddot{H}
+ \dddot{H}\big)f''(R) + 36\big(4H\dot{H} + \ddot{H}\big)^2f'''(R)\bigg]
\label{pressure}
\end{eqnarray}
respectively, where $f(R)$ is the deviation of $F(R)$ gravity from the 
Einstein gravity, that is $F(R) = R + f(R)$. Thus, the effective energy-momentum tensor (EMT)
depends on the form of $F(R)$, as expected. Therefore, different 
forms of $F(R)$ will lead to different evolution of the Hubble parameter. We will use such effective EMT 
of $F(R)$ gravity to realize the cosmological evolution of the Universe.\\

In the present context, we are interested to study a unified scenario of a non-singular bounce and dark energy epoch, 
in particular, from a non-singular bounce to a deceleration epoch and from
a deceleration epoch to a late time acceleration era. The bouncing scenario 
requires a violation of energy condition(s), which we incorporate here through a  modified gravity theory in the form of a higher curvature 
gravity model. This is motivated from the fact that the spacetime curvature becomes large during the bounce and thus it is natural 
to generalize the Einstein-Hilbert action by adding higher order curvature terms in the gravitational action. Such higher curvature terms 
may also naturally arise from the diffeomorphism property of the action. Furthermore, with the help of higher curvature 
gravity theories, the unification of inflation and dark energy epochs has been studied earlier by some of our authors \cite{Nojiri:2019fft,
Nojiri:2020wmh}, thereby we hope that higher curvature theories may have a significant role also in unifying the bounce with dark energy epoch. 
In particular, here we consider a Chern-Simons (CS) corrected 
F(R) theory in a four dimensional spacetime model. The importance of Chern-Simons Lagrangian from various perspectives 
has been reported in \cite{Bajardi:2021hya}. Here we would like to mention that in four dimensional spacetime, a special case of
Lanczos-Lovelock theory, namely the Gauss-Bonnet (GB) theory becomes a total surface term and thus has no contributions in the field equations. 
The scenario however becomes different in higher dimensional spacetime 
where the GB term affects the field equations non-trivially \cite{Bajardi:2021hya}. Coming back to our present model, i.e the 
Chern-Simons corrected F(R) model in four dimensional FRW spacetime, the Chern-Simons term shows no contributions in the background Friedmann equations, 
and thus the background evolution is entirely controlled by the vacuum F(R) term (see \ref{basic4}). This is in contrast to 
\cite{Bajardi:2021hya} where the authors considered an AdS invariant Chern-Simons Lagrangian in $five~dimensional$ spacetime, which can be 
recovered from the Lovelock theory, and found the corresponding cosmological as well as spherically symmetric solutions. 
On contrary, in our present scenario (which is a four dimensional spacetime model), the 
effective energy density and pressure arises from the higher curvature F(R) degrees of freedom drive the background evolution. However the  
vacuum F(R) theory is known to predict a large tensor-to-scalar ratio in the bouncing cosmology, which is not consistent with the Planck data. 
Thus in order to reduce the tensor-to-scalar ratio, we consider the Chern-Simons term along with the F(R) model. Although, the Chern-Simons 
term does not affect the background FRW equations, it indeed modifies the perturbation evolution considerably. Moreover it has been shown earlier that 
the Chern-Simons term helps to reduce the tensor-to-scalar ratio  in an $inflationary$ spacetime \cite{Odintsov:2019mlf}. 
This motivates us to consider the Chern-Simons corrected F(R) model to explore the unification of a non-singular bounce to the dark energy epoch, 
with a hope that due to the effect of the CS term, the tensor-to-scalar ratio gets reduced (and fit within the Planck constraint) 
compared to that of in the vacuum F(R) case.

\section{Background evolution}\label{sec-background}
As mentioned earlier, we are interested in getting an unified cosmological picture of a non-singular bounce to a late time 
dark energy epoch. In this regard, the background scale factor present in the FRW metric is considered as \cite{Odintsov:2016tar},
\begin{eqnarray}
 a(t) = \left[1 + a_0\left(\frac{t}{t_0}\right)^2\right]^n\exp{\left[\frac{1}{(\alpha-1)}\left(\frac{t_s - t}{t_0}\right)^{1-\alpha}\right]}~,
 \label{scale factor1}
\end{eqnarray}
where $a_0$, $n$, $\alpha$ are positive valued dimensionless parameters, 
while the other ones like $t_s$ and $t_0$ have the dimensions of time. The parameter 
$t_0$ is taken to scale the cosmic time in billion years, so we take $t_0 = 1\mathrm{By}$ (the $\mathrm{By}$ stands for 'billion years' 
throughout the paper) and consequently, the scale factor becomes, 
\begin{eqnarray}
 a(t) = \left[1 + a_0t^2\right]^n\times \exp{\left[\frac{1}{(\alpha-1)}\left(t_s - t\right)^{1-\alpha}\right]} = a_1(t)\times a_2(t)~(\mathrm{say})~~.
 \label{scale factor2}
\end{eqnarray}
The scale factor is taken as a product of two factors- $a_1(t)$ and $a_2(t)$ respectively, where the factor $a_2(t)$ is motivated 
in getting a viable dark energy epoch at late time. Actually 
$a(t) = a_1(t)$ is sufficient for getting a non-singular bouncing universe where the bounce occurs at $t = 0$. However 
at late expanding phase of the universe, the scale factor $a_1(t)$ goes as $\sim t^{2n}$, which, in turn, does not lead to a viable dark energy model 
according to the Planck results. Thereby, in order to get a bounce along with a viable dark energy epoch, we consider the scale factor 
as of \ref{scale factor2} where $a_1(t)$ is multiplied by $a_2(t)$. 
We will show that the presence of $a_2(t)$ does not harm the bouncing character of the universe, however it slightly shifts the bouncing time 
from $t = 0$ to a negative time and moreover the scale factor (\ref{scale factor2}) leads to an asymmetric bounce scenario (as $a(t) \neq a(-t)$). 
In particular, due to the exponential term, the $a_2(t)$ seems to have negligible effects at negative values of $t$ 
i.e during the contracting universe; however it shows considerable effects during the expanding phase, which, along with $a_1(t)$, leads to a 
viable dark energy epoch. We will come back to this point in details at some stage.\\
The scale factor of \ref{scale factor2} immediately leads to the Hubble parameter ($H = \dot{a}/a$) and the Ricci scalar 
($R(t) = 12H^2 + 6\dot{H}$) as (an overdot represents $\frac{d}{dt}$ of the respective quantity),
\begin{eqnarray}
 H(t) = \frac{1}{a}\frac{da}{dt} = \frac{2a_0nt}{\left(1 + a_0t^2\right)} + \frac{1}{\left(t_s - t\right)^{\alpha}}
 \label{Hubble parameter}
\end{eqnarray}
and
\begin{eqnarray}
 R(t) = \frac{12a_0n}{\left(1 + a_0t^2\right)^2}\left\{1 - a_0t^2\left(1-4n\right)\right\} 
 + \frac{12}{\left(t_s - t\right)^{2\alpha}} + \frac{6\alpha}{\left(t_s - t\right)^{1+\alpha}} + \frac{48a_0nt}{\left(1 + a_0t^2\right)
 \left(t_s - t\right)^{\alpha}}
 \label{ricci scalar}
\end{eqnarray}
respectively. \ref{Hubble parameter} refers different types of finite time singularity at $t = t_s$ (see \cite{Nojiri:2005sx} for different 
types of finite time future singularity), where the singularity structure depends 
on the value of $\alpha$. In particular,
\begin{itemize}
 \item For $\alpha > 1$, a Type-I singularity appears at $t = t_s$, i.e the scale factor, the effective energy density ($\rho_\mathrm{eff}$) 
 and the effective pressure ($p_\mathrm{eff}$) simultaneously diverge at $t = t_s$.  
 
 \item For $0 < \alpha < 1$, a Type-III singularity occurs at $t = t_s$, i.e the scale factor tends to a finite value, while the $\rho_\mathrm{eff}$ and 
 $p_\mathrm{eff}$ diverge at $t = t_s$.
 
 \item For $-1 < \alpha < 0$, a Type-II singularity appears at $t = t_s$. In this case, the scale factor and $\rho_\mathrm{eff}$ tend to a finite value, 
 while $p_\mathrm{eff}$ diverges at $t = t_s$.
 
 \item For $\alpha < -1$ and non-integer, 
 a Type-IV singularity appears at $t = t_s$, in which case, the scale factor, $\rho_\mathrm{eff}$ and $p_\mathrm{eff}$ 
 tend to a finite value $t = t_s$, however the higher derivatives of the Hubble parameter diverge at the singularity point.
\end{itemize}

Therefore the finite time singularity at $t = t_s$ is almost inevitable in the present context. Thus in order to describe a singularity free universe's 
evolution up-to the present epoch ($\approx 13.5\mathrm{By}$), we consider the parameter $t_s$ to be greater than the present age of the universe, 
i.e $t_s > t_p \approx 13.5\mathrm{By}$. Therefore with this condition, we may argue that the Hubble parameter of \ref{Hubble parameter} 
describes a singularity free cosmological evolution up-to $t \gtrsim t_p$. During the cosmic time $t \gg t_p$ : either the universe will hit to the 
finite time singularity at $t = t_s$ (predicted by the present model) 
or possibly some more fundamental theory will govern that regime by which the finite time singularity can be avoided. 

Coming back to our present model, the Hubble horizon at distant past gets the evolution as $|1/aH| \sim |t|^{1-2n}$ which, for $n < 1/2$ (which is 
indeed compatible with the Planck results in regard to the observables like spectral index, tensor to scalar ratio- as we will demonstrate in 
\ref{sec-perturbation}), 
diverges to infinity at $t \rightarrow -\infty$. This indicates that the primordial perturbations generate in the deep contracting era when 
all the perturbation modes are within the Hubble radius. Such generation era of the primordial perturbation modes is similar to that of in the matter 
bounce scenario. However the matter bounce model leads to a deceleration phase at late expanding phase and thus is 
not consistent with dark energy model, unlike to the present bounce scenario which indeed leads to a dark energy era 
at the present epoch.\\ 
In regard to the background evolution at late contracting era- the scale factor, Hubble parameter and the Ricci scalar 
have the following expressions:
\begin{eqnarray}
 a(t) \approx a_0^nt^{2n}~~~~~~~~,~~~~~~~~H(t) \approx \frac{2n}{t}~~~~~~~~\mathrm{and}~~~~~~~~R(t) \approx -\frac{12n(1-4n)}{t^2}~~.
 \label{late contracting expressions}
\end{eqnarray}
With these expressions, the F(R) gravitational \ref{basic4} turns out to be, 
\begin{eqnarray}
 \left(\frac{2}{1 - 4n}\right)R^2\frac{d^2F}{dR^2} - \left(\frac{1 - 2n}{1 - 4n}\right)R\frac{dF}{dR} + F(R) = 0~~,
 \label{gravitational equation-late contracting}
\end{eqnarray}
on solving which, we get the form of F(R) at late contracting era as,
\begin{eqnarray}
 F(R) = R_0\left[\left(\frac{R}{R_0}\right)^{\rho} + \left(\frac{R}{R_0}\right)^{\delta}\right]
 \label{FR solution-late contracting}
\end{eqnarray}
where $R_0$ is a constant, and the exponents $\rho$, $\delta$ have the following forms (in terms of $n$),
\begin{eqnarray}
 \rho = \frac{1}{4}\left[3 - 2n - \sqrt{1 + 4n\left(5+n\right)}\right]~~~~~,~~~~~
 \delta = \frac{1}{4}\left[3 - 2n + \sqrt{1 + 4n\left(5+n\right)}\right]
 \label{rho-delta}
\end{eqnarray}
respectively. From \ref{FR solution-late contracting}, we get the expression of $F'(R)$ which proves to be useful 
to investigate the stability of the primordial perturbations in the present context,
\begin{eqnarray}
F'(R) = \frac{\rho}{\left(R/R_0\right)^{1-\rho}} + \delta \left(R/R_0\right)^{\delta-1}
 \label{F-prime-R}
\end{eqnarray}
In order to have a clear understand, we give the plots of $\rho$ and $\delta$ with respect to $n$ in \ref{plot-rho-delta} which
clearly demonstrates that $\delta$ remains positive for all possible values of $n$, however in the case of $\rho$ - it 
starts from a positive value at $n = 0$ and gets a zero crossing from positive to negative values at $n = 1/4$. Depending on the choices whether 
$n < 1/4$ or $n > 1/4$, we get two different physical pictures in regard to the sign of $F'(R)$. The demonstration goes as follows:

\begin{figure}[!h]
\begin{center}
 \centering
 \includegraphics[scale=0.70]{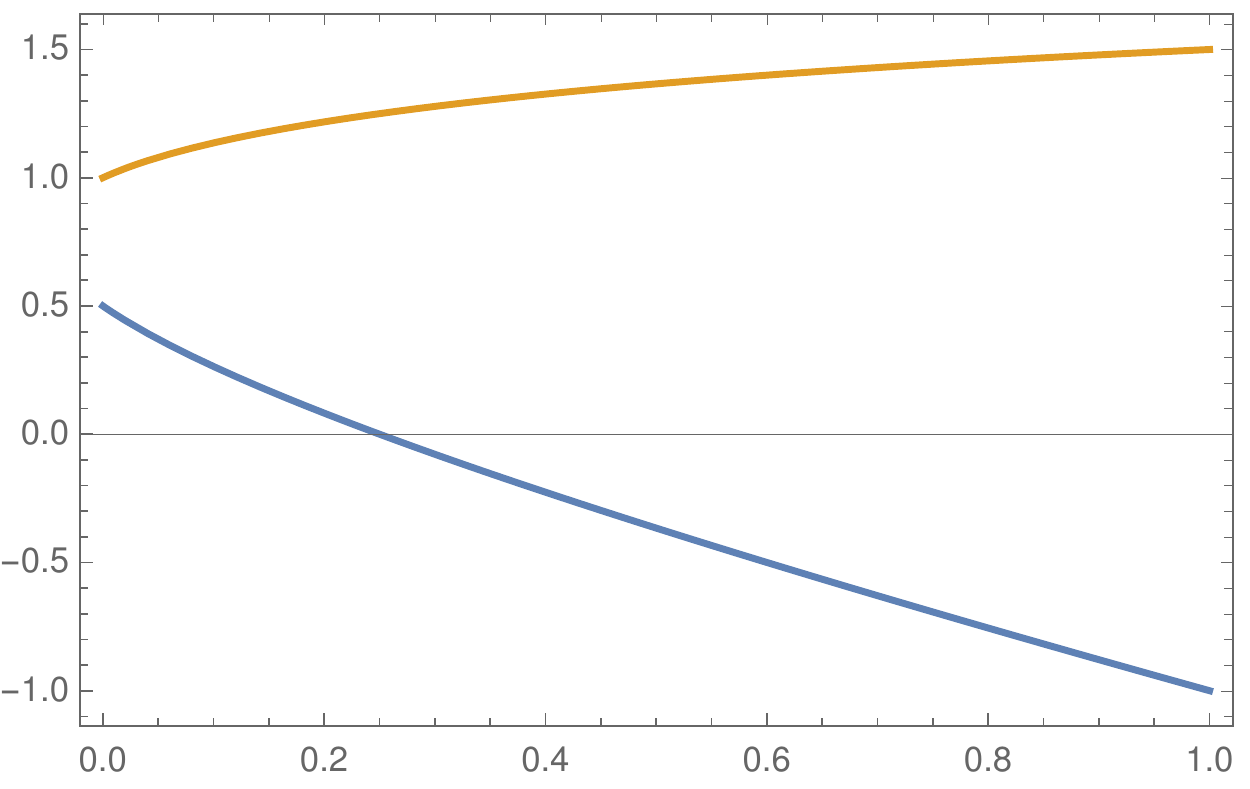}
 \caption{{\underline{Blue Curve}}: $\rho$ vs. $n$ and {\underline{Yellow Curve}}: $\delta$ vs. $n$ ; as per the \ref{rho-delta}.}
 \label{plot-rho-delta}
\end{center}
\end{figure}

\begin{itemize}
 \item $The~case~n > 1/4$: In this regime, $\rho$ is negative. Therefore at late contracting era (at $t \rightarrow -\infty$) where 
 $R \rightarrow 0$, the dominant term of $F'(R)$ is the first term of \ref{F-prime-R} which makes the $F'(R)$ negative at distant past. This in turn 
 yields the instability of the primordial perturbations in the deep contracting regime when the perturbation modes generate inside the Hubble radius.
 
 \item $The~case~n < 1/4$: In this regime, $0 < \rho < 1$ and $\delta > 1$. Thereby both the terms present in the $F'(R)$ are positive 
 and thus $F'(R) > 0$ in the deep sub-Hubble regime. This makes the primordial perturbations stable at their generation era.
\end{itemize}

Based on the above discussions, we consider $n < 1/4$ in the present work. However for $n < 1/4$, the Ricci scalar at distant past, from 
\ref{late contracting expressions}, becomes negative and thus in order to get real valued function of $F(R)$, we consider the constant 
$R_0$ (appeared in the solution of $F(R)$ in \ref{FR solution-late contracting}) to be negative as well. Here it deserves mentioning that 
\ref{late contracting expressions} is valid at distant past, from which it seems that 
the Ricci scalar is negative, only at late contracting era. In particular, by using the expression of $R(t)$ in \ref{ricci scalar} which is valid 
for entire cosmic time, we will show that the $R(t)$ indeed has a zero crossing from negative to 
positive values at the contracting era. In particular, the zero crossing (from negative to positive values) 
of $R(t)$ occurs before the instant of bounce, and after that zero crossing, the Ricci scalar remains positive throughout the cosmic time. Before moving 
to the full evolution of $R(t)$, let us demonstrate the bouncing behaviour of the $a(t)$ of \ref{scale factor2}.

\subsection{Realization of a non-singular asymmetric bounce}\label{sec-bounce-realization}
In this section, we will show that the scale factor (\ref{scale factor2}) allows a non-singular bounce at a finite time. The universe's evolution 
in a general bouncing cosmology consists of two eras, an era of contraction where the Hubble parameter is negative and an era of expansion having 
positive Hubble parameter. In particular, the bounce phenomena is defined by the conditions $H(t) = 0$ and $\dot{H} > 0$ respectively. 
To realize this in the present context, we borrow the expression of $H(t)$ from \ref{Hubble parameter}, i.e
\begin{eqnarray}
 H(t) = \frac{2a_0nt}{\left(1 + a_0t^2\right)} + \frac{1}{\left(t_s - t\right)^{\alpha}}~~.
 \label{Hubble parameter-bounce1}
\end{eqnarray}
As the parameters $a_0$, $n$ and $\alpha$ are positive, the Hubble parameter during $t > 0$ remains positive. However during negative time, 
i.e for $t < 0$, the first term of \ref{Hubble parameter-bounce1} becomes negative while the second term remains positive, 
thus there is a possibility to have $H(t) = 0$ and $\dot{H} > 0$ at some negative $t$. Let us check it more explicitly.

For $t < 0$, we can write $t = -|t|$ and \ref{Hubble parameter-bounce1} can be expressed as,
\begin{eqnarray}
 H(t) = -\frac{2a_0n|t|}{\left(1 + a_0|t|^2\right)} + \frac{1}{\left(t_s + |t|\right)^{\alpha}} = -H_1(t) + H_2(t)~(\mathrm{say})~~.
 \label{Hubble parameter-bounce2}
\end{eqnarray}
The term $H_1(t)$ starts from the value zero at $t \rightarrow -\infty$ and reaches to zero at $t = 0$, with an extremum (in particular, a maximum) 
at an intermediate stage of $-\infty < t < 0$. However the second term $H_2(t)$ starts from the value zero at $t \rightarrow -\infty$ and reaches 
to $1/t_s^{\alpha}$ at $t = 0$, with a monotonic increasing behaviour during $-\infty < t < 0$. 
Furthermore both the $H_1(t)$ and $H_2(t)$ increase at 
$t \rightarrow -\infty$ with their respective rate of increasing are given by: 
\begin{eqnarray}
 \frac{dH_1}{dt}\bigg|_{t\rightarrow -\infty} \sim 1/|t|^2~~~~~~~,~~~~~~~ \frac{dH_2}{dt}\bigg|_{t\rightarrow -\infty} = 1/|t|^{(1+\alpha)}\nonumber
\end{eqnarray}
respectively, i.e $H_1(t)$ increases at a faster rate compared to that of $H_2(t)$ at $t \rightarrow -\infty$ for $\alpha > 1$. 
Here it may be mentioned that the condition $\alpha > 1$ is also related to the positivity of the Ricci scalar, as 
we will establish in \ref{model parameter constraint-1}, and thus $\alpha > 1$ is well justified in the present context. 
Such evolutions of $H_1(t)$ and $H_2(t)$ during $t \leq 0$ are given in the following \ref{Table-1}, 
which are essential to realize a bouncing behaviour of the Hubble parameter:\\

 \begin{table}[t]
  \centering
  \begin{tabular}{|c|c|}
   \hline 
  $Evolution~of~H_1(t)$  & $Evolution~of~H_2(t)$\\
  \hline
   \hline\\
   $H_1(t) = 0$ at $t\rightarrow -\infty$ & $H_2(t) = 0$ at $t\rightarrow -\infty$ \\
   \hline\\
   $H_1(t) = 0$ at $t = 0$ & $H_2(t) = 1/t_s^{\alpha}$ (i.e positive) at $t = 0$ \\
   \hline\\
   $H_1(t)$ has a maximum at $-\infty < t < 0$ & $H_2(t)$ monotonically increases during $-\infty < t < 0$ \\
   \hline
   $dH_1/dt\bigg|_{t\rightarrow -\infty} \sim 1/|t|^2$ & $dH_2/dt\bigg|_{t\rightarrow -\infty} \sim 1/|t|^{(1+\alpha)}$ \\
   \hline
   \hline
  \end{tabular}%
  \caption{Comparison between the evolutions of $H_1(t)$ and $H_2(t)$ during $t \leq 0$.}
  \label{Table-1}
 \end{table}
 
Therefore at $t \rightarrow -\infty$ : both the $H_1(t)$ and $H_2(t)$ start from the value zero with an increasing behaviour, however their rate 
of increasing are different to each other, in particular $H_1(t)$ increases at a faster rate compared to that of $H_2(t)$. Moreover 
at $t = 0$ : the term $H_2(t)$ is positive while $H_1(t) = 0$, i.e $H_2(t)$ becomes larger than $H_1(t)$. These arguments clearly indicate that there 
exits a negative finite $t$, say $t = -\tau$ with $\tau$ being positive, for which the following statements of $H(t)$ hold true:
\begin{itemize}
 \item $H_1(t) > H_2(t)$ or equivalently $H(t) < 0$ during $t < -\tau$,
 
 \item $H_1(t) = H_2(t)$ or equivalently $H(t) = 0$ at $t = -\tau$,
 
 \item $H_1(t) < H_2(t)$ or equivalently $H(t) > 0$ during $t > -\tau$.
\end{itemize}
Therefore, $t = -\tau$ (with $\tau > 0$) is the instant when the bounce occurs, and can be determined by the condition $H(-\tau) = 0$, i.e,
\begin{eqnarray}
 \frac{2a_0n\tau}{\left(1 + a_0\tau^2\right)} = \frac{1}{\left(t_s + \tau\right)^{\alpha}}~~.
 \label{bounce-instant}
\end{eqnarray}
In order to see whether the above equation has a real solution (for $\tau$), we need to put some definite values of the parameters, as \ref{bounce-instant} 
may not be solved in a closed form. In the following two subsections, we will determine the constraints on the model parameters from various requirements 
(like - to get a late time accelerating stage, the effective EoS parameter 
at present epoch matches with the Planck results \cite{Aghanim:2018eyx} etc.). However for instance, let us choose some 
specific values of the parameters which indeed match with the constraints that we will determine after \ref{Planck1}, 
in particular, we consider : $a_0 = 0.30$, 
$n = 0.185$, $t_s = 30$ and $\alpha = 4/3$, for which the algebraic \ref{bounce-instant} yields $\tau \approx 0.09$ (recall the bounce occurs at 
$t = -\tau$). Furthermore, we would like to mention that the scale factor remains positive at the instant of bounce, 
particularly the aforementioned parametric regime leads to  $a(-\tau) \approx 2.62$, i.e the bounce in the present context 
is indeed a non-singular bounce.\\

In regard to the time evolution of the Ricci scalar, \ref{late contracting expressions} clearly indicates that $R(t)$ behaves as $\sim -1/t^2$ 
at distant past, i.e the Ricci scalar starts from $0^{-}$ at $t \rightarrow -\infty$. However at the instant of bounce, the $R(t)$ becomes positive, 
due to the reason that the Hubble parameter vanishes and its derivative is positive at the bounce point. Therefore the Ricci scalar 
must undergo a zero crossing from negative to positive value before the bounce occurs. At this stage, 
we require that after that zero crossing, the Ricci scalar 
remains to be positive throughout the cosmic time, which can be realized by a more stronger condition that the Ricci scalar has to be positive 
during the expanding phase of the universe, in particular,
\begin{eqnarray}
 R(t > -\tau) > 0~~,
 \label{requirement1}
\end{eqnarray}
where we may recall that $t = -\tau$ is the instant of bounce. 
Here we will determine the constraints on the model parameters such that the above requirement holds true. For this purpose, we borrow the expression 
of $R(t)$ from \ref{ricci scalar} as,
\begin{eqnarray}
 R(t) = \frac{12a_0n}{\left(1 + a_0t^2\right)^2}\left\{1 - a_0t^2\left(1-4n\right)\right\} 
 + \frac{12}{\left(t_s - t\right)^{2\alpha}} + \frac{6\alpha}{\left(t_s - t\right)^{1+\alpha}} + \frac{48a_0nt}{\left(1 + a_0t^2\right)
 \left(t_s - t\right)^{\alpha}}
 \label{ricci scalar1}
\end{eqnarray}
The only term (present in the above expression) due to which the Ricci scalar $may$ acquire negative values during the $expanding~phase$ is given by 
the second term $a_0t^2\left(1 - 4n\right)$, in particular during $a_0t^2\left(1 - 4n\right) > 1$, the terms within the curly bracket 
provide negative contributions to $R(t)$. Thereby we consider the duration $\frac{1}{a_0(1-4n)} < t^2 < t_s^2$ where 
the Ricci scalar can be expressed as,
\begin{eqnarray}
 R(t) = \frac{12a_0n}{\left(1 + a_0t^2\right)^2}\bigg\{1 + \frac{4}{t_s^{\alpha}}t - a_0t^2\left(1 - 4n\right) + \frac{4a_0}{t_s^{\alpha}}t^3\bigg\} 
 + \frac{12}{t_s^{2\alpha}} + \frac{6\alpha}{t_s^{1+\alpha}}~~.
 \label{ricci scalar2}
\end{eqnarray}
The last two terms in the above expression contribute positive values to the Ricci scalar and 
thereby in order to determine the constraint(s) (on the model parameters) 
corresponds to the requirement $R(t > -\tau) > 0$, we can only consider the terms that are within the curly bracket of \ref{ricci scalar2}. Let us denote 
it by $\widetilde{R}(t)$, i.e
\begin{eqnarray}
 \widetilde{R}(t) = \bigg\{1 + \frac{4}{t_s^{\alpha}}t - a_0t^2\left(1 - 4n\right) + \frac{4a_0}{t_s^{\alpha}}t^3\bigg\}
 \label{tilde-R}
\end{eqnarray}
It is clear that the condition $\widetilde{R} > 0$ during the expanding phase in turn leads to our requirement given in \ref{requirement1}. 
One can check that for $n < 1/4$ (which is indeed our consideration to make the primordial perturbations stable, as discussed 
earlier after \ref{F-prime-R}), the $\widetilde{R}(t)$ during the expanding phase becomes positive if the model parameters 
satisfy the following relations:
\begin{eqnarray}
 \alpha > 1~~~~~~~~\mathrm{and}~~~~~~~
 \frac{\sqrt{a_0}\left(1 - 4n\right)}{8\sqrt{3}} < \frac{1}{t_s^{\alpha}} < \frac{\sqrt{a_0}\left(1 - 4n\right)}{4\sqrt{3}}
 \label{model parameter constraint-1}
\end{eqnarray}
respectively. We determine the above constraints by finding the minimum of $\widetilde{R}(t)$ (from its derivative with respect to $t$) 
and use the condition that $\widetilde{R}_{min} > 0$ : which is indeed a necessary condition to make a function positive valued. Thus as a whole, 
\ref{model parameter constraint-1} confirms that the Ricci scalar remains positive after its zero crossing which, in fact, 
occurs before the instant of bounce. Hence we stick to these parameter constraints throughout the paper. Furthermore, here we would like to mention 
that one of the above constraints $\alpha > 1$ leads to a Type-I singularity at $t = t_s$, i.e the Hubble parameter as well as 
the $\rho_\mathrm{eff}$ and $p_\mathrm{eff}$ diverge at $t = t_s$, as mentioned after \ref{ricci scalar}. However, since $t_s \gtrsim t_p$, the 
present model satisfactorily describes a singularity free cosmological evolution up-to $t \gtrsim t_p$ with $t_p \approx 13.5\mathrm{By}$ 
being the present age of the universe.

\subsection{Acceleration and deceleration stages of the expanding universe}

The acceleration factor of the universe is given by $\ddot{a}/a = \dot{H} + H^2$ which, from \ref{Hubble parameter}, turns out the be,
\begin{eqnarray}
 \frac{\ddot{a}}{a} = \frac{2a_0n\left\{1 - a_0t^2(1 - 2n)\right\}}{\left(1 + a_0t^2\right)^2} 
 + \frac{\alpha}{\left(t_s - t\right)^{1+\alpha}} + \frac{4a_0nt}{\left(1 + a_0t^2\right)\left(t_s - t\right)^{\alpha}} 
 + \frac{1}{\left(t_s - t\right)^{2\alpha}}~~.
 \label{acceleration-1}
\end{eqnarray}
To understand the acceleration or deceleration stages of the universe, we need to give the plot of the above expression of $\ddot{a}/a$, 
which in turn requires the values of the model parameters present in \ref{acceleration-1}. However before moving to such 
quantitative description and the corresponding plot of $\ddot{a}/a$, first we 
want to analyze that how much information(s) of $\ddot{a}/a$ can obtain $qualitatively$ from \ref{acceleration-1}.\\     

\begin{figure*}[t!]
\centering
\subfloat[\label{Fig_3a}]{\includegraphics[scale=0.38]{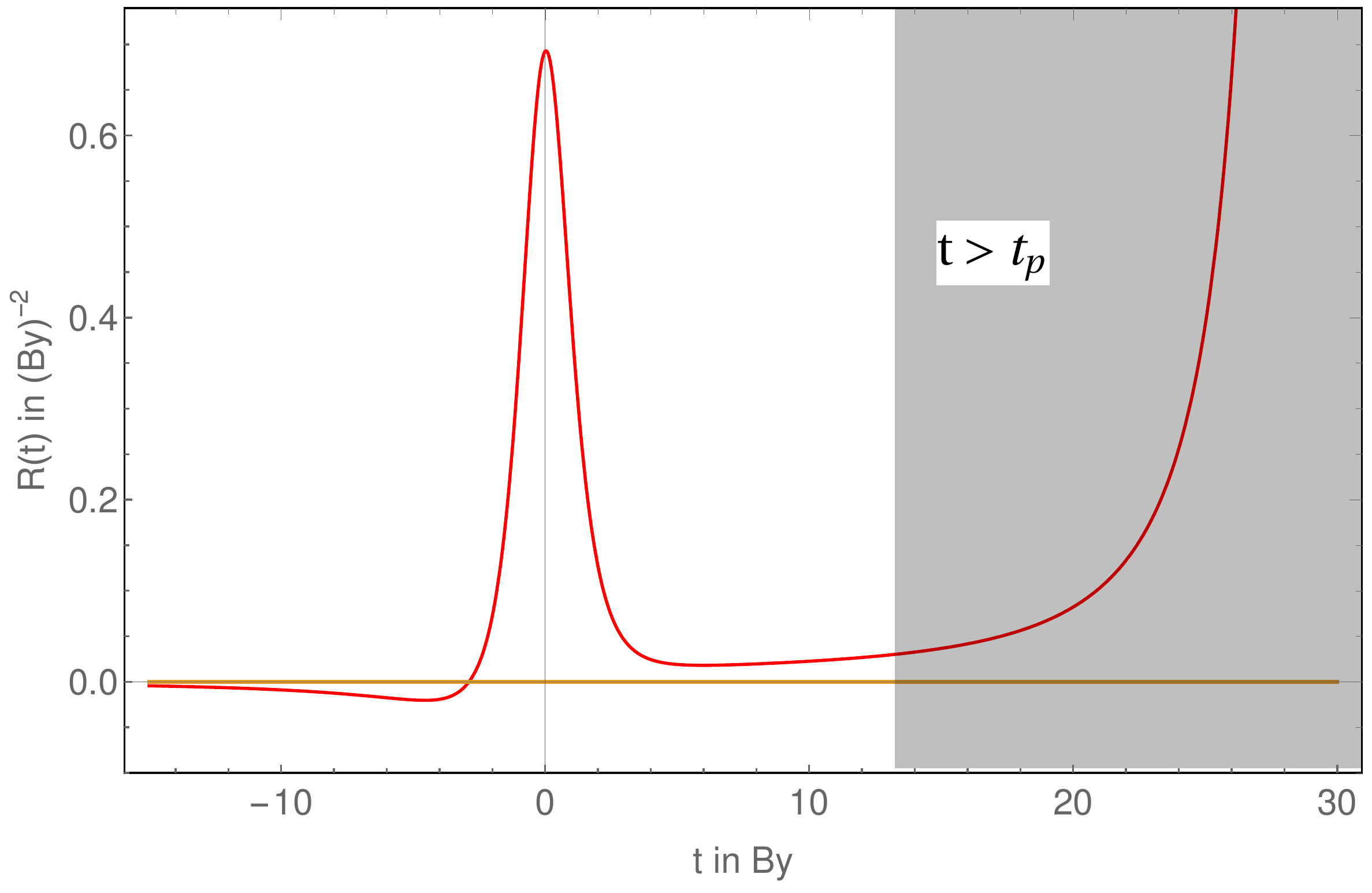}}~~
\subfloat[\label{Fig_3b}]{\includegraphics[scale=0.35]{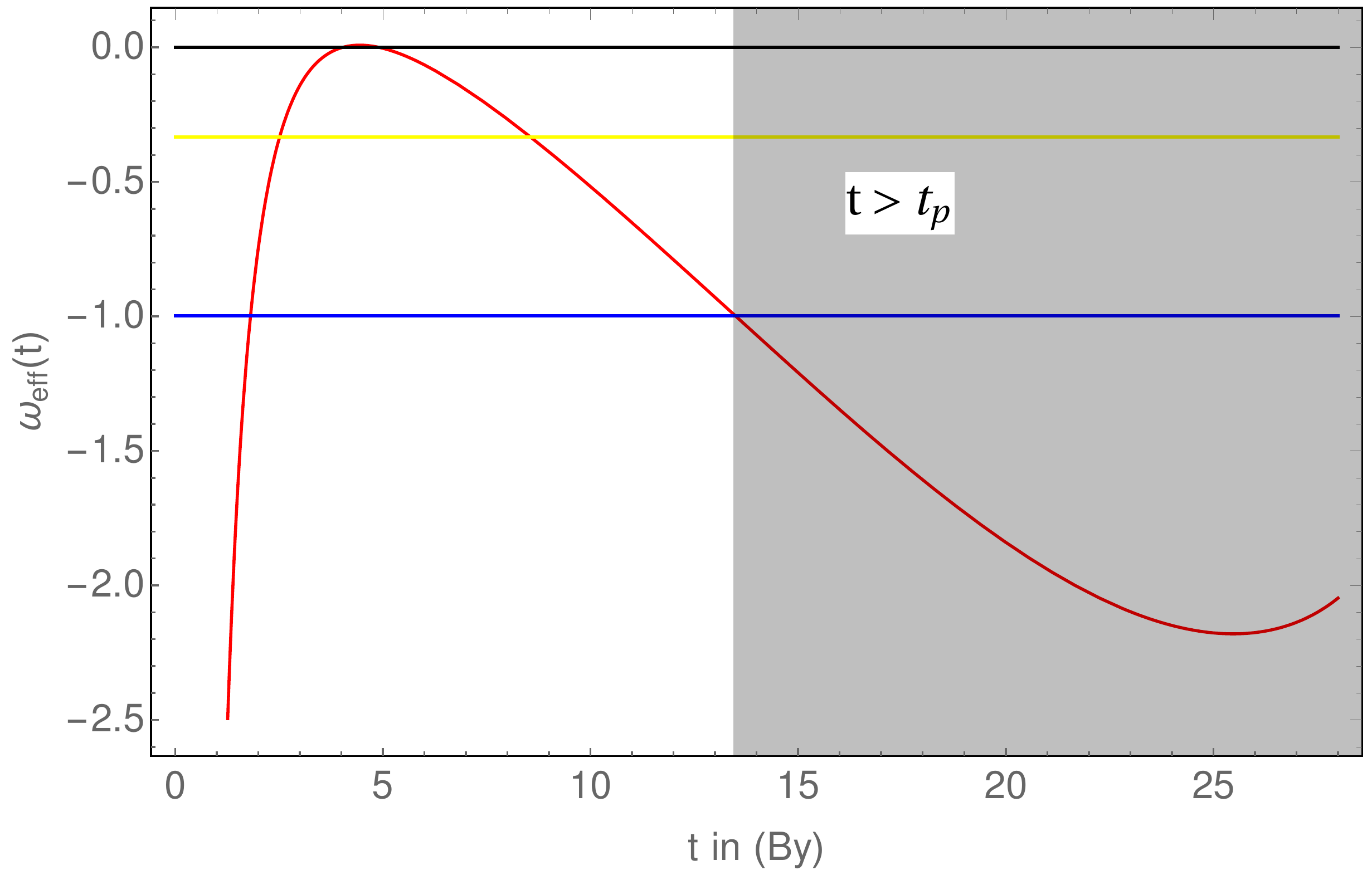}}
\caption{The above figure depicts the time evolution of (a) the Ricci scalar $R(t)$ and 
(b) the EoS parameter $\omega_\mathrm{eff}(t)$. Both the plots correspond to $n = 0.185$, $\alpha = 4/3$, $t_s = 30$ 
and $a_0 = 0.30$. Moreover the shaded region in the plots correspond to the cosmic time larger than the present age of the universe, 
i.e $t > t_p \approx 13.5\mathrm{By}$. In the right plot, the black, yellow and blue curve correspond to $\omega_\mathrm{eff} = 0,-\frac{1}{3},-0.997$ 
respectively. The curve $\omega_\mathrm{eff} = 0$ helps to investigate whether the present model exhibits a matter-like dominated epoch during some regime 
of cosmic time, the curve $\omega_\mathrm{eff} = -1/3$ is to demonstrate the accelerating or decelerating stages of the universe and the 
$\omega_\mathrm{eff} = -0.997$ curve reveals that the effective EoS of the present model matches with the Planck results at the present epoch i.e at 
$t = 13.5\mathrm{By}$.}
\label{plot-Ricci-EoS}
\end{figure*}

It is evident that near $t \approx 0$, $\frac{\ddot{a}}{a} \approx 2a_0n + \frac{\alpha}{t_s^{1+\alpha}} + \frac{1}{t_s^{2\alpha}}$, 
i.e $\ddot{a}$ is positive. 
This is however expected, because $t \approx 0$ is the bouncing regime where, 
due to the fact that $\dot{H} > 0$ near the bounce, the universe undergoes through an accelerating stage. However, as $t$ increases particularly 
during $t^2 > \frac{1}{a_0(1 - 2n)}$, the first term of \ref{acceleration-1} becomes negative and hence the universe may expand through 
a decelerating phase. As $t$ increases further, the terms containing $1/(t_s - t)$ starts to grow at a faster rate compared to the other terms 
(since $\alpha$ is positive) and possibly $\ddot{a}$ becomes positive, i.e the universe may transit from a decelerating phase to an accelerating one. 
The transition of the universe 
from acceleration to deceleration or vice-versa can be described by $\ddot{a} = 0$ which, due to \ref{acceleration-1}, is expressed as,
\begin{eqnarray}
 \frac{2a_0n\left\{1 - a_0t^2(1 - 2n)\right\}}{\left(1 + a_0t^2\right)^2} 
 + \frac{\alpha}{\left(t_s - t\right)^{1+\alpha}} + \frac{4a_0nt}{\left(1 + a_0t^2\right)\left(t_s - t\right)^{\alpha}} 
 + \frac{1}{\left(t_s - t\right)^{2\alpha}} = 0~~.
 \label{acceleration-2}
\end{eqnarray}
The above algebraic equation of $t$ may not be solved in a closed form, however based on the above arguments, we consider two different 
regimes of the cosmic time to understand the transition from acceleration to deceleration (or vice-versa) of the expanding universe.
\begin{itemize}
 \item During,  $\frac{1}{a_0\left(1 - 2n\right)} < t^2 \ll t_s^2$ : In this regime of $t$, \ref{acceleration-2} can be written as
 \begin{eqnarray}
  \frac{4a_0n}{t_s^{\alpha}}t^3 - 2a_0n\left(1 - 2n\right)t^2 + 2n = 0
  \label{acceleration-3}
 \end{eqnarray}
where $t_s - t \approx t_s$ is considered. Here we need to investigate whether the above algebraic equation of $t$ has a solution in the regime 
$\frac{1}{a_0\left(1 - 2n\right)} < t^2 \ll t_s^2$; thus we may consider the solution ansatz 
as $t_1 = \frac{1}{\sqrt{a_0\left(1-2n\right)}}\left(1 + \delta\right)$, with $t_1$ being the root of \ref{acceleration-3} and $\delta < 1$. 
Plugging back the ansatz into \ref{acceleration-3} and retaining up-to the first order in $\delta$ yields 
$\delta = \frac{a_0}{t_s^{\alpha}\left(a_0(1-2n)\right)^{3/2}}$ and consequently $t_1$ is given by
\begin{eqnarray}
 t_1 = \frac{1}{\sqrt{a_0\left(1-2n\right)}}\left(1 + \frac{a_0}{t_s^{\alpha}\left(a_0(1-2n)\right)^{3/2}}\right)~.
 \label{t1}
\end{eqnarray}
Thereby, we may argue that \ref{acceleration-3} contains a root in the aforesaid regime for at $t = t_1$. 
This along with the fact that the universe passes through an acceleration near the bounce, indicate 
that during $\frac{1}{a_0\left(1 - 2n\right)} < t^2 \ll t_s^2$ the universe makes a transition from the accelerating phase to a decelerating one. 
Later we will show that this is indeed the case when we give the plot of the full effective EoS in \ref{Fig_3b}. 

\item During,  $\frac{1}{a_0\left(1 - 2n\right)} \ll t^2 < t_s^2$ : In this regime of $t$, \ref{acceleration-2} can be expressed as,
\begin{eqnarray}
\left(\frac{1}{t_s^{2\alpha}}\right)t^2 + \frac{2\alpha}{t_s}\left(1 - 2n\right)t - \left(1 - 2n\right) = 0
\label{acceleration-4}
\end{eqnarray}
where the term $\frac{1}{\left(t_s - t\right)^{2\alpha}}$ is considered to be the dominant piece compared to the others. According to the 
``Descartes rule of signs'', \ref{acceleration-4} must contain one positive and one negative real root for $t$. The positive root (say at $t_2$) 
is obtained as,
\begin{eqnarray}
 t_2 = \alpha t_s^{2\alpha - 1}\left(1 - 2n\right)
 \bigg\{\sqrt{1 + \frac{1}{\alpha^2\left(1 - 2n\right)t_s^{2\alpha - 2}}} - 1\bigg\} \simeq \frac{t_s}{2\alpha}~,
 \label{acceleration-solution}
\end{eqnarray}
which, due to $\alpha > 1$, is less than $t_s$. The existence of the above root depicts that 
during $\frac{1}{a_0\left(1 - 2n\right)} \ll t^2 < t_s^2$, the universe makes another transition from the intermediate 
decelerating phase to an accelerating one and continues the expansion in that accelerating stage. Finally at $t = t_s$, 
$\ddot{a}$ diverges, which is expected due to the occurrence of the Type-I singularity at $t = t_s$ in the present context.\\ 

\end{itemize}
Thereby \ref{acceleration-1} has two positive real roots for $t$ during the regime $\frac{1}{a_0\left(1 - 2n\right)} < t^2 \ll t_s^2$ and 
$\frac{1}{a_0\left(1 - 2n\right)} \ll t^2 < t_s^2$ respectively. 
As a whole, the picture is following: (1) the universe undergoes through an accelerating stage near the bounce during the expanding era, (2) 
as $t$ increases, particularly during $\frac{1}{a_0\left(1 - 2n\right)} < t^2 \ll t_s^2$, the universe gets a transition from the accelerating phase to a 
decelerating phase and (3) with further increases of $t$, the universe makes a second and final 
transition from the intermediate decelerating stage to an accelerating one.  The second transition from $\ddot{a} < 0$ to $\ddot{a} > 0$ 
is identified with the $late~time~acceleration$ epoch of the universe. Therefore we require $t_2 \lesssim t_p$, where $t_2$ is the instant 
of the second transition and recall, $t_p$ represents the present age of the universe. In particular, \ref{acceleration-4} has the solution at,
\begin{eqnarray}
 t_2 = \alpha t_s^{2\alpha - 1}\left(1 - 2n\right)\bigg\{\sqrt{1 + \frac{1}{\alpha^2\left(1 - 2n\right)t_s^{2\alpha - 2}}} - 1\bigg\} \simeq \frac{t_s}{2\alpha}
 \label{acceleration-solution}
\end{eqnarray}
where in getting the second equality, we expand the terms within the square root binomially (as $\alpha > 1$ and also $t_s \gtrsim t_p$) 
and retain upto the first order. Therefore the above solution of $t_2$ along with the requirement $t_2 \lesssim t_p$ put a constraint 
on the model parameters as,
\begin{eqnarray}
 t_s \lesssim 2\alpha t_p~~.
 \label{constraint2}
\end{eqnarray}
Here we need to recall that the present model predicts a Type-I singularity at $t = t_s$ (since $\alpha > 1$) 
and thus in order to get a singularity free cosmological evolution upto the cosmic time $t \gtrsim t_p$, the parameter $t_s$ satisfies $t_s > t_p$. 
Combining this condition with \ref{constraint2}, we get the allowed range of $t_s$ as,
\begin{eqnarray}
 t_p < t_s \lesssim 2\alpha t_p~~.
 \label{constraint3}
\end{eqnarray}
The EoS parameter of the dark energy epoch is defined as $\omega_\mathrm{eff}(t) = -1-\frac{2\dot{H}}{3H^2}$, where $H(t)$ is shown 
in \ref{Hubble parameter}. With this expression of $\omega_\mathrm{eff}$, we confront the model with the latest Planck+SNe+BAO results which put 
a constraint on the dark energy EoS parameter as \cite{Aghanim:2018eyx}
\begin{eqnarray}
 \omega_\mathrm{eff}(t_p) = -0.957 \pm 0.080
 \label{Planck1}
\end{eqnarray}
with $t_p \approx 13.5\mathrm{By}$ being the present age of the universe. Thereby we choose the model parameters in such a way that the above constraint 
on $\omega_\mathrm{eff}(t_p)$ holds true.\\

As a whole, we have four parameters in our hand: $n$, $a_0$, $t_s$ and $\alpha$. Below is the list of their constraints that we found earlier from 
various requirements,
\begin{itemize}
 \item $C1$ : The parameter $n$ is constrained by $n < 1/4$ in order to make the primordial perturbations stable at the deep sub-Hubble radius 
 in the contracting era.
 
 \item $C2$ : $t_s$ is larger than the present age of the universe, i.e $t_s > t_p \approx 13.5\mathrm{By}$ to describe a singularity free evolution 
 of the universe upto the cosmic time $t \gtrsim t_p$.
 
 \item $C3$ : $t_s \lesssim 2\alpha t_p$ in order to have an accelerating stage of the present universe. This along with the previous condition 
 $C2$ lead to $t_p < t_s \lesssim 2\alpha t_p$.
 
 \item $C4$ : In regard to the parameters $\alpha$ and $a_0$, they are found to be constrained as $\alpha > 1$ and 
 $\frac{\sqrt{a_0}\left(1 - 4n\right)}{8\sqrt{3}} < \frac{1}{t_s^{\alpha}} < \frac{\sqrt{a_0}\left(1 - 4n\right)}{4\sqrt{3}}$. These make 
 the Ricci scalar positive after its zero crossing at the contracting era. In particular, the zero crossing (from negative to positive values) of the 
 Ricci scalar occurs before the instant of the bounce.
 
 \item $C5$ : $\omega_\mathrm{eff}(t_p) = -0.957 \pm 0.080$, to confront the theoretical expectations of the dark energy EoS with the Planck+SNe+BAO 
 results.
 \end{itemize}
 
\begin{figure}[!h]
\begin{center}
 \centering
 \includegraphics[scale=0.45]{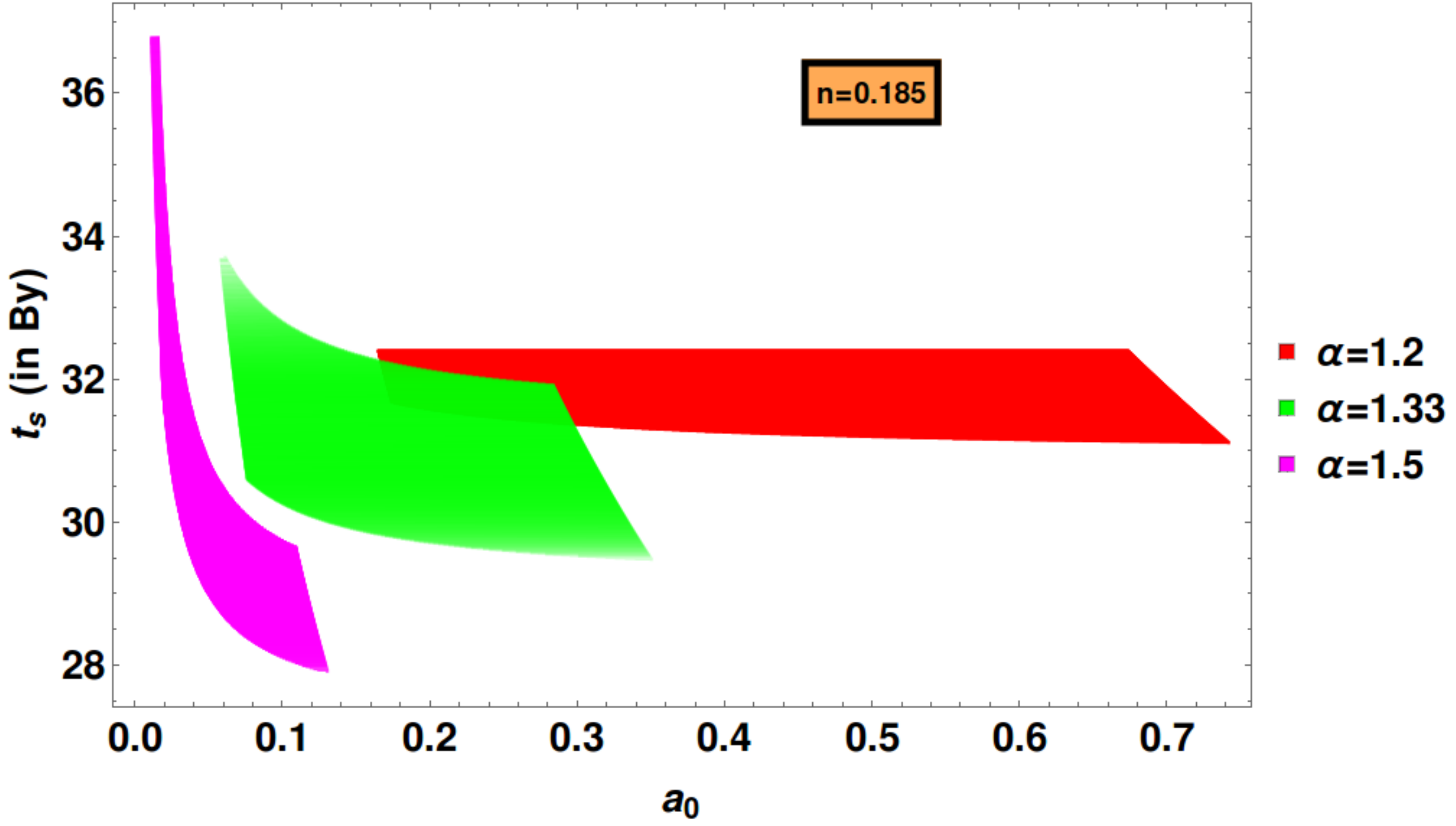}
 \caption{The contour regions of $t_s$ (in the unit of By.) vs. $a_0$ for three different set of 
 values of $\alpha$ and $n$, in particular -- $(\alpha,n) = (1.2,0.185)$, $(\alpha,n) = (\frac{4}{3},0.185)$ and 
 $(\alpha,n) = (1.5,0.185)$ respectively.}
 \label{plot-contour}
\end{center}
\end{figure}

 In order to better understand the above constraints on the model parameters, we give contour plots in \ref{plot-contour} depicting the allowed 
 regions of the parameters. The aforesaid constraints from $C1$ to $C5$ are taken care in 
 \ref{plot-contour} which actually demonstrates the variation of $t_s$ vs. $a_0$ for three different set of 
 values of $\alpha$ and $n$, in particular we take -- $(\alpha,n) = (1.2,0.185)$, $(\alpha,n) = (\frac{4}{3},0.185)$ and 
 $(\alpha,n) = (1.5,0.185)$ respectively. The value $n = 0.185$ is motivated due to the fact that this certain value of $n$ 
 leads to the $central~value$ of the spectral index of scalar curvature perturbation, as we will show 
 in \ref{sec-perturbation} during the analysis the cosmological perturbation.

 \begin{figure*}[t!]
\centering
\subfloat[\label{Fig_2a}]{\includegraphics[scale=0.35]{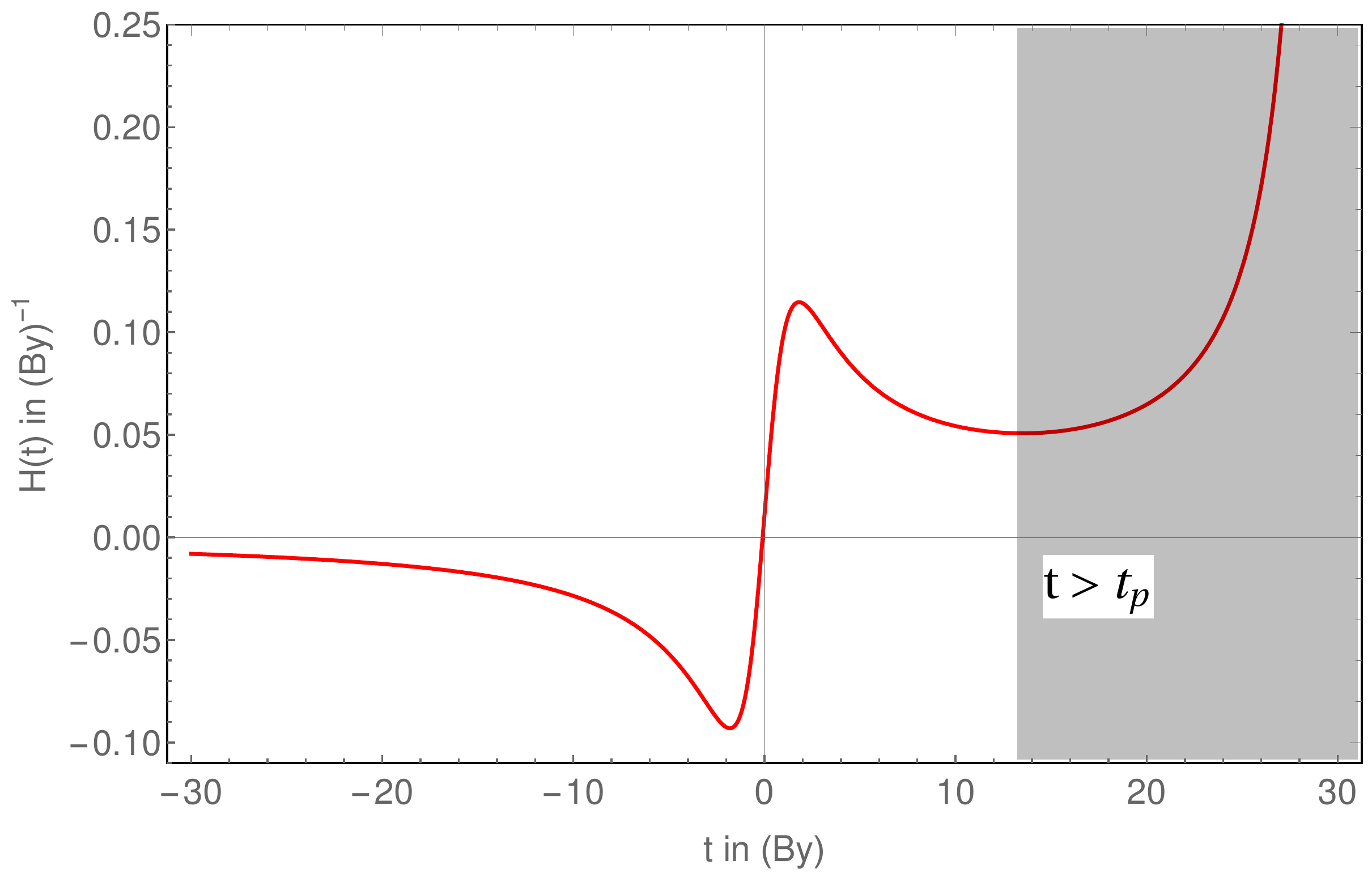}}~~
\subfloat[\label{Fig_2b}]{\includegraphics[scale=0.32]{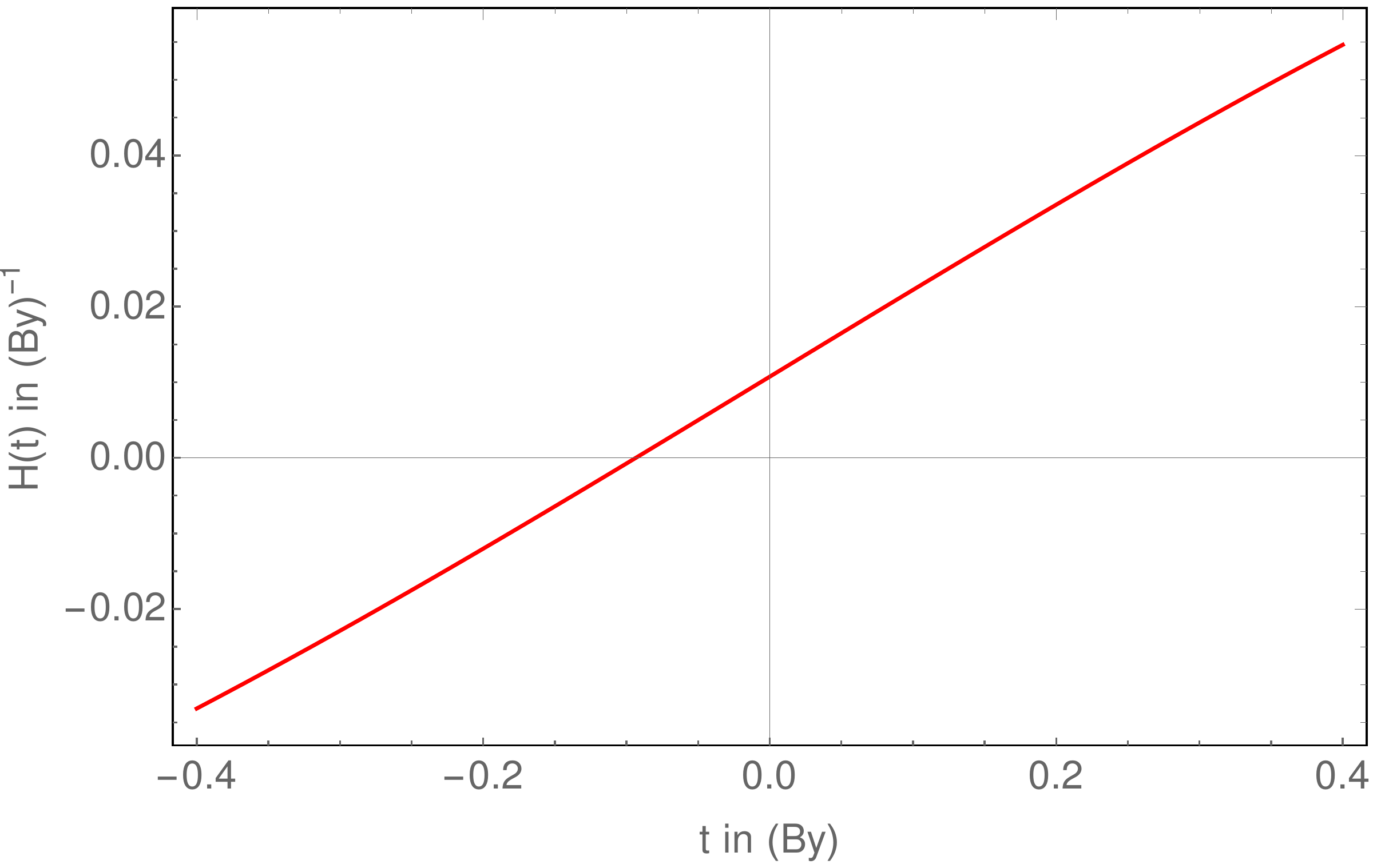}}
\caption{The above figure depicts the time evolution of (a) the Hubble parameter $H(t)$ and 
(b) the zoomed-in version of $H(t)$ near the bounce. Both the plots correspond to $n = 0.185$, $\alpha = 4/3$, $t_s = 30$ 
and $a_0 = 0.30$. Moreover the shaded region in the left plot corresponds to the cosmic time larger than the present age of the universe, 
i.e $t > t_p \approx 13.5\mathrm{By}$.}
\label{plot-Hubble}
\end{figure*}

 Keeping the parameter constraints in mind, we further give the plots of the background 
 $H(t)$, $R(t)$ and $\omega_\mathrm{eff}(t)$ (with respect to cosmic time) 
 by using \ref{Hubble parameter} and \ref{ricci scalar}, see \ref{plot-Hubble} and \ref{plot-Ricci-EoS}. 
 The parameter values are considered as $n = 0.185$, $\alpha = 4/3$ 
 and $t_s = 30$ in getting the plots, which are within the green region of \ref{plot-contour} and thus are allowed in the present context. 
 With such values of 
 $n$, $\alpha$, $t_s$, the condition $C4$ gives $0.08 < a_0 < 0.33$ while the $C5$ leads to $a_0 > 0.12$. Combining these two, one may take 
 $0.12 < a_0 < 0.33$ to satisfy both the $C4$ and $C5$, and consequently we consider $a_0 = 0.30$ in the plots. Actually this specific value of $a_0$ 
 (along with $n = 0.185$, $\alpha = 4/3$ and $t_s = 30$) leads to $\omega_\mathrm{eff}(t_p) = -0.997$, which is in agreement with \cite{Elizalde:2010ts} 
 where the authors considered an exponential F(R) model to explain the dark energy model, (here it may be mentioned that 
 such value of current $\omega_\mathrm{eff}$ is also consistent with the holographic dark energy model, see \cite{Nojiri:2021iko}). 
 \ref{Fig_2a} clearly depicts that the Hubble parameter becomes zero and shows an increasing behaviour with time (i.e $\dot{H} > 0$) near 
 $t \approx 0$, which indicates the instant of a non-singular bounce. In particular, $t \approx -0.09\mathrm{By}$ is the time when the bounce occurs 
 (see \ref{Fig_2b}), which is consistent with the arguments of \ref{sec-bounce-realization} where we realized a non-singular bounce at some negative $t$. 
 In regard to the evolution of the Ricci scalar, \ref{Fig_3a} demonstrates that $R(t)$ starts from $0^{-}$ at asymptotic past. The negative 
 values of $R(t)$ during the asymptotic past is due to \ref{late contracting expressions} with $n < 1/4$. However 
 as the cosmic time increases, the Ricci scalar gets a zero crossing from negative to positive values at the contracting era. In particular, such 
 zero crossing of $R(t)$ happens before the bounce occurs and after that zero crossing the Ricci scalar seems to be positive throughout the 
 cosmic time. This, in fact, is in agreement with the constraint $C4$ as mentioned above. Here we would like to mention that 
 both the Hubble parameter and the Ricci scalar diverge at $t = t_s = 30\mathrm{By}$ (recall we consider $t_s = 30\mathrm{By}$ in the plots), 
 which in turn refers to a Type-I singularity at $t = t_s$. However as evident from the figures that the occurrence of the Type-I singularity 
 is far away from the present age of the universe and thus we argue that the present model satisfactorily describes a singularity free cosmological 
 evolution of the universe upto $t \gtrsim t_p$. Coming to the evolution of the effective EoS parameter, 
 the red curve of \ref{Fig_3b} represents the $w_{eff}(t)$ for the present model while the yellow one of the same is for the constant 
 value $-\frac{1}{3}$ (we will keep the yellow graph to investigate the accelerating or decelerating era of the Universe). 
 \ref{Fig_3b} clearly demonstrates that near the bounce i.e. near $t \approx 0$, the EoS parameter diverges from the negative side, 
 however this is expected because at the bounce the Hubble 
 parameter itself becomes zero and in turn makes the $w_{eff} = -1 - \frac{2\dot{H}}{3H^2} \rightarrow -\infty$. Then after the bounce, 
 $w_{eff}$ crosses the value $-\frac{1}{3}$ leading to a transition 
 from a bounce to a decelerating phase of the Universe, and during the deceleration epoch, the EoS parameter becomes zero during an epoch 
 indicating a matter-like dominated Universe. The deceleration era continues till $t \approx 8.5\mathrm{By}$ 
 when the EoS parameter again crosses the value $-\frac{1}{3}$ and thus the 
 Universe transits from a stage of deceleration to a stage of acceleration which, in turn, is identified with the present dark energy epoch. 
 Therefore, the present model may provide an unified scenario of certain cosmological epochs from bounce to late-time acceleration followed 
 by a matter-like dominated epoch in the intermediate regime. Moreover, the present value of the dark energy EoS parameter seems to be 
 $\omega_\mathrm{eff}(t_p) \simeq -0.997$ from the \ref{Fig_3b} $\big($the blue curve is for the constant value $-0.997$, that coincides with the 
 red one at $t = t_p \approx 13.5\mathrm{By}\big)$, which is indeed consistent with the Planck-2018+SNe+BAO results \cite{Aghanim:2018eyx}.
 
\begin{figure*}[t!]
\centering
\subfloat[\label{Fig_5a}]{\includegraphics[scale=0.65]{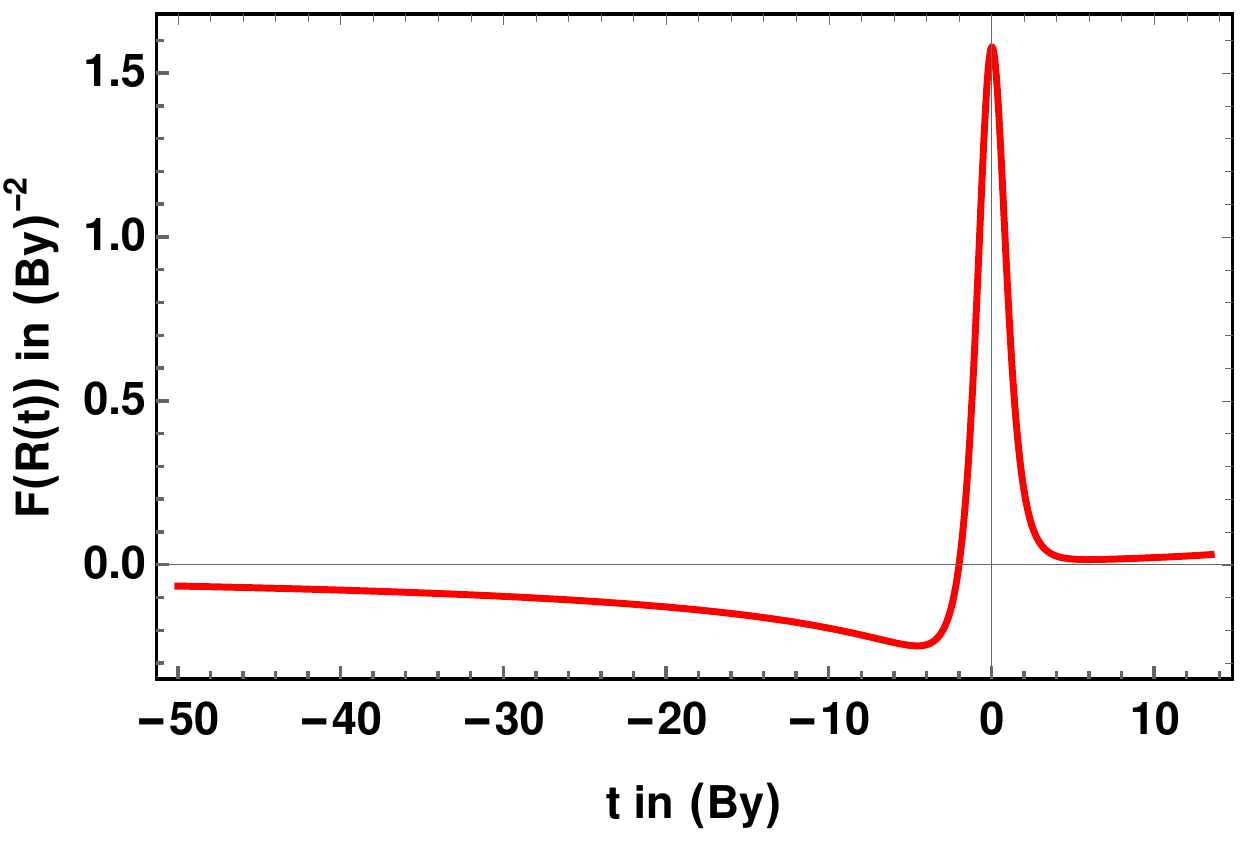}}~~
\subfloat[\label{Fig_5b}]{\includegraphics[scale=0.45]{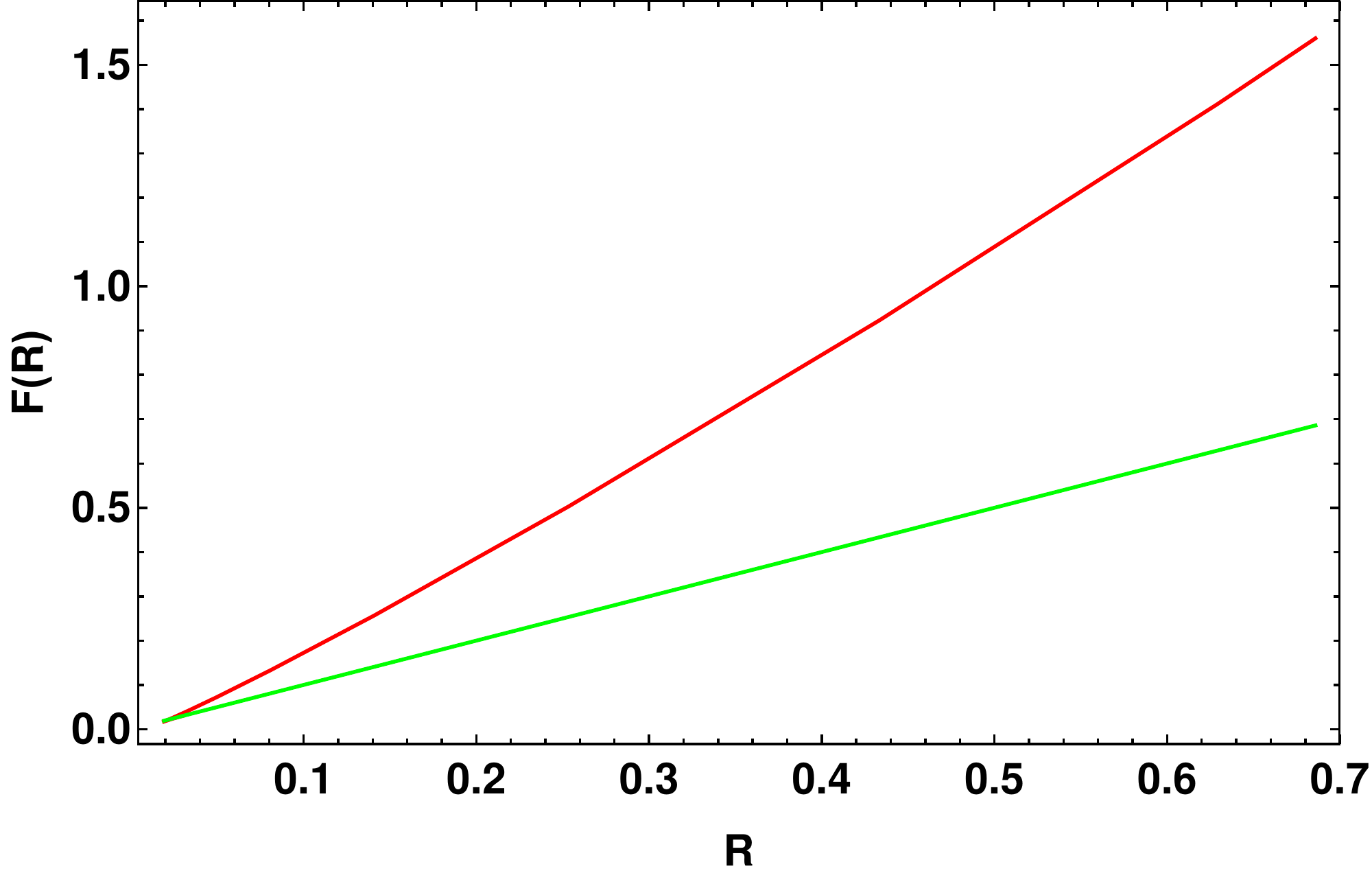}}
\caption{The above figure depicts- (a) the time evolution of $F(R(t))$ and 
(b) the $F(R)$ vs. $R$. The initial condition during this numerical analysis is considered to be 
$F(R(t)) = \left(R(t)/R_0\right)^{\rho} + \left(R(t)/R_0\right)^{\delta}$ with $R_0 = -1\mathrm{(By)}^{-2}$, 
recall the parameter $R_0$ is negative, as discussed after \ref{F-prime-R}. 
The right plot is obtained from the parametric plot of $F(R(t))$ and $R(t)$. 
Both the plots correspond to $n = 0.185$, $\alpha = 4/3$, $t_s = 30$ and $a_0 = 0.30$.}
\label{plot-F-solution}
\end{figure*}

 The remaining task is to determine the form of $F(R)$ from the gravitational \ref{basic4}, which 
 leads to such background cosmological evolution of the universe. In accordance the form of $H(t)$ in \ref{Hubble parameter}, 
 the F(R) gravitational equation may 
 not be solved analytically and thus we will solve it numerically. For this purpose, \ref{basic4} is re-written in terms of cosmic time as,
 \begin{eqnarray}
  H(t)\dot{R}(t)\frac{d^F}{dt^2} - \left\{\dot{H}\dot{R} + H^2\dot{R} + H\ddot{R}\right\}\frac{dF}{dt} + \frac{1}{6}\dot{R}^2(t)F(t) = 0~,
  \label{F-solution1}
 \end{eqnarray}
where $F(t) = F(R(t))$. Using the form of $H(t)$ from \ref{Hubble parameter} along with the expression $R(t) = 12H^2 + 6\dot{H}$, we numerically solve 
\ref{F-solution1} for $F = F(t)$ during a wide range of cosmic time, see \ref{Fig_5a}. The initial condition of this numerical analysis 
is considered to be $F(R(t)) = \left(R(t)/R_0\right)^{\rho} + \left(R(t)/R_0\right)^{\delta}$, i.e the analytic form of $F(R(t))$ 
during the late contracting era is taken as the initial condition of the numerical solution of \ref{F-solution1}. Such numerical solution of 
$F= F(t)$ along with the expression of $R = R(t)$ (see \ref{ricci scalar}) lead to the form of $F(R)$ (by using ``parametric plot'' in MATHEMETICA), 
see \ref{Fig_5b}. Actually the form of  $F(R)$ is demonstrated by the red curve, while the green one represents the Einstein gravity. 
\ref{Fig_5b} clearly depicts that the F(R) in the present context matches with the Einstein gravity as the Ricci scalar approaches to the present value, 
while the F(R) seems to deviate from the usual Einstein gravity, when the scalar curvature takes larger and larger values. 
It is evident that $F'(R)$ is positive, which, as we will show in 
\ref{sec-perturbation}, is connected to the stability of the primordial perturbation near the bounce; moreover $F'(R) > 0$ 
also indicates that the model is free from the Ostrogradsky instability.\\

Here we would like to mention that in regard to the background evolution, the effective EoS parameter at distant past is given by 
$\omega_\mathrm{eff} = -1+\frac{1}{3n}$ which is indeed less than unity due to the aforementioned range of $n$ that makes 
the observable quantities viable with the Planck results. In effect, the anisotropic energy density 
grows as $a^{-6}$ during the contracting era and thus the background evolution in the contracting stage becomes unstable to the growth of 
anisotropies, which is known as BKL instability \cite{new1}. Thereby like many other bounce models, 
the present model is suffered from the BKL instability. However on contrary, 
in the ekpyrotic bounce scenario, the bouncer field decays more faster than the anisotropic energy density, by which the BKL instability 
can be resolved \cite{Cai:2013kja,Cai:2013vm,Erickson:2003zm,Garfinkle:2008ei}. 
Therefore it may be an interesting avenue to unify an ekpyrotic bounce with a dark energy epoch in some appropriate modified 
theories of gravity, which we expect to address in future work. Moreover the Chern-Simons F(R) gravity 
can be extended by axion DM, as considered in \cite{Odintsov:2020nwm,Nojiri:2020pqr} in the context of inflationary background spacetime. 
The possible effects of DM may also be included in the current unified scenario of bounce and dark energy epochs, which will be considered elsewhere.

 \section{Cosmological perturbation}\label{sec-perturbation}
 
 In this section, we perform the cosmological perturbation of the background spacetime in the present context and consequently determine 
 various observable quantities like scalar spectral index, tensor to scalar ratio etc. In a bouncing scenario, the Hubble parameter vanishes 
 and thus the comoving Hubble radius (defined by $r_h = 1/(aH)$) diverges at the instant of bounce. However the asymptotic behaviour of the Hubble radius 
 may be different in different bounce models, which qualitatively differ various bounce model(s) in regard to the generation era 
 of the primordial perturbation. In some bounce models (for example, the matter bounce scenario), the Hubble radius monotonically 
 increases with time at late contracting era and asymptotically diverges at $t \rightarrow -\infty$, due to which the perturbation modes 
 in such bounce models generate at deep contracting era where all the perturbation modes lie within the horizon. On other hand, there may exist 
 some bounce models (see \cite{Odintsov:2020zct}) where the Hubble radius decreases with time and asymptotically goes to zero at $t \rightarrow -\infty$; 
 in such scenario, the primordial perturbation generate near the bounce where the Hubble horizon has an infinite size to contain all the perturbation 
 modes within it.
 
 In the present context, the scale factor at late contracting era behaves as $a(t) \sim |t|^{2n}$ and thus the Hubble radius goes as 
 $r_h \sim |t|^{1-2n}$. Here we need to recall that $n < 1/4$ (see the aforementioned condition $C1$), due to which the Hubble radius diverges at 
 $t \rightarrow -\infty$. This makes the generation era of the primordial perturbation at the early contracting stage within the deep 
 sub-Hubble radius. We would like to mention that the scale factor is asymmetric with respect to the bounce point, in 
 particular, unlike to the fact that the Hubble radius diverges at $t \rightarrow -\infty$, it monotonically decreases with cosmic time 
 at the late stage of the expanding era. Actually, such decreasing behaviour of the Hubble radius ensures a dark energy epoch of the present universe.  
 
 \subsection{Scalar Perturbations}

The scalar perturbation of FRW background metric is defined as
follows,
\begin{align}
ds^2 = -(1 + 2\Psi)dt^2 + a(t)^2(1 - 2\Psi)\delta_{ij}dx^{i}dx^{j}\, ,
\label{sp1}
\end{align}
where $\Psi(t,\vec{x})$ denotes the scalar perturbation. Here we are working in the comoving gauge, in which case 
the curvature perturbation ($\mathcal{R}(t,\vec{x})$) becomes identical to the $\Psi(t,\vec{x})$ and thus we can proceed with the perturbation 
variable $\Psi(t,\vec{x})$. The perturbed action up to $\Psi^2$ order is \cite{Hwang:2005hb,Noh:2001ia,Hwang:2002fp},
\begin{align}
\delta S_{\psi} = \int dt d^3\vec{x} a(t) z(t)^2\left[\dot{\Psi}^2
 - \frac{1}{a^2}\left(\partial_i\Psi\right)^2\right]\, ,
\label{sp2}
\end{align}
where $z^2(t)$, in the present context of Chern-Simons corrected F(R) gravity theory, has the following expression \cite{Hwang:2005hb},
\begin{align}
z^2(t) = \frac{a^2(t)}{\kappa^2\left(H(t) + \frac{1}{2F'(R)}\frac{dF'(R)}{dt}\right)^2} \bigg\{\frac{3}{2F'(R)}\left(\frac{dF'(R)}{dt}\right)^2\bigg\}\, .
\label{sp3}
\end{align}
As it can be observed from the above form of $z(t)$, the Chern-Simons (CS) term does not affect the scalar perturbation; 
due to the reason that it is not possible to form a scalar energy momentum tensor nor vector or symmetric tensor, which contains 
the Levi-Civita tensor and scalar derivatives only \cite{Hwang:2005hb}. However the CS term indeed affects the tensor type perturbation, which we will 
demonstrate in the next section. Coming back to \ref{sp2}, it is evident 
that the speed of the scalar perturbation waves (or the sound speed) is unity, which indicates the 
absence of the superluminal modes from the present model, or equivalently we may argue that the model is free from the gradient 
instability. The stability of the scalar perturbations is ensured from the condition $z^2(t) > 0$ which along with \ref{sp3} leads to 
$F'(R) > 0$. We have the expression of $F'(R)$ at contracting stage in \ref{F-prime-R}, 
which is indeed positive due to $n < 1/4$ -- this indicates the stability of the scalar perturbation during 
the contracting stage at where the perturbation modes generate. Moreover the numerical solution of $F(R)$ in \ref{plot-F-solution} clearly indicates that 
the $F'(R)$ is positive during wide range of cosmic time, which makes the scalar perturbations stable in the present context.

Here we are interested to determine various observable quantities like the scalar spectral index and tensor-to-scalar ratio which are eventually 
evaluated at the time of horizon crossing of the large scale modes. Due to the fact that the Hubble radius diverges at $t \rightarrow -\infty$, the large 
scale modes cross the horizon at deep contracting stage (later we will explicitly calculate the horizon crossing instant of the large scale modes) 
where the Hubble parameter and $F(R)$ follow \ref{late contracting expressions} and 
\ref{FR solution-late contracting} respectively. Thereby using such expressions of $H(t)$ and $F(R)$, we determine various terms present in
the expression of $z(t)$ (see \ref{sp3}) as,
\begin{align}
\frac{a(t)}{\left(H(t) + \frac{1}{2F'(R)}\frac{dF'(R)}{dt}\right)}
= \frac{a_0^n\left(12n(4n-1)\right)^{n+1/2}}{\left(-R\right)^{n+1/2}}
\left\{2n + \frac{(1-\rho)\left[1 + \frac{\delta(\delta-1)}{\rho(\rho-1)}
\left(\frac{R}{R_0}\right)^{\delta-\rho}\right]}
{\left[1 + \frac{\delta}{\rho}\left(\frac{R}{R_0}\right)^{\delta-\rho}\right]}\right\}^{-1}\, ,
\nonumber
\end{align}
and
\begin{align}
\frac{3}{2F'(R)}\left(\frac{dF'(R)}{dt}\right)^2
= |R_0|\left(\frac{R}{R_0}\right)^{\rho} \left\{\frac{\rho(1-\rho)^2
\left[1 + \frac{\delta(\delta-1)}{\rho(\rho-1)}
\left(\frac{R}{R_0}\right)^{\delta-\rho}\right]^2} {2n(1-4n)\left[1
+ \frac{\delta}{\rho}\left(\frac{R}{R_0}\right)^{\delta-\rho}\right]}\right\}\, .
\nonumber
\end{align}
where, recall, $R(t)$ is negative at late contracting era due to $n < 1/4$ and thus $\left(-R\right)^{n+1/2}$ present in the previous expression 
is real. Consequently $z(t)$ takes the following form,
\begin{eqnarray}
z(t) = \left\{\frac{a_0^n|R_0|^n\left[12n(1-4n)\right]^n}{\kappa\left(R/R_0\right)^{n+1/2-\rho/2}}\right\}\times\left(\frac{P(R)}{Q(R)}\right)
\label{sp4}
\end{eqnarray}
where $P(R)$ and $Q(R)$ are defined as follows,
\begin{align}
P(R) = \left\{\frac{\sqrt{\rho}(1-\rho)
\left[1 + \frac{\delta(\delta-1)}{\rho(\rho-1)}
\left(\frac{R}{R_0}\right)^{\delta-\rho}\right]}{\left[1
+ \frac{\delta}{\rho}\left(\frac{R}{R_0}\right)^{\delta-\rho}\right]^{1/2}}\right\}~~,
\label{P}
\end{align}
and
\begin{align}
Q(R) = \left\{2n + \frac{(1-\rho)\left[1 + \frac{\delta(\delta-1)}{\rho(\rho-1)}
\left(\frac{R}{R_0}\right)^{\delta-\rho}\right]}
{\left[1 + \frac{\delta}{\rho}\left(\frac{R}{R_0}\right)^{\delta-\rho}\right]}\right\}~~.
\label{Q}
\end{align}

\ref{sp2} clearly indicates that $\Psi(t,\vec{x})$ is not
canonically normalized and to this end we introduce the well-known
Mukhanov-Sasaki variable as $v = z\Re$ ($= z\Psi$ as we are
working in the comoving gauge). The corresponding Fourier mode of
the Mukhanov-Sasaki variable satisfies,
\begin{align}
\frac{d^2v_k}{d\eta^2} + \left(k^2 - \frac{1}{z(\eta)}\frac{d^2z}{d\eta^2}\right)v_k(\eta)
= 0 \, ,
\label{sp5}
\end{align}
where $\eta = \int dt/a(t)$ is the conformal time and $v_k(\eta)$
is the Fourier transformed variable of $v(t,\vec{x})$ for the
$k$ -th mode. \ref{sp5} may not be solved analytically in general, as $z(\eta)$ depends on the background evolution. However the equation 
can be solved at late contracting era, as we now demonstrate. The conformal time $\eta$ is related to the cosmic time as, 
$\eta(t) = \frac{1}{a_0^n\left(1-2n\right)}t^{1-2n}$ for $n \neq 1/2$; since the parameter $n$ is constrained to be less than $1/4$, we can safely 
work with this expression of $\eta(t)$. Using the $\eta = \eta(t)$, we can express the Ricci scalar as a
function of the conformal time,
\begin{eqnarray}
R(\eta) = \frac{1}{\eta^{2/(1-2n)}}\left\{\frac{12n(1-4n)}{|R_0|\left[a_0^n(1-2n)\right]^{2/(1-2n)}}\right\} \propto \frac{1}{\eta^{2/(1-2n)}}
\label{sp6}
\end{eqnarray}
Having this in mind, we can express $z(\eta)$ from \ref{sp4} in terms of $\eta$ as follows,
\begin{align}
z(\eta) \propto \left(\frac{P(\eta)}{Q(\eta)}\right)\times\eta^{\frac{2n+1-\rho}{1-2n}}
\label{sp67}
\end{align}
where $P(\eta) = P(R(\eta))$ and $Q(\eta) = Q(R(\eta))$, with $P(R)$, $Q(R)$ are given in \ref{P}, \ref{Q} respectively. The 
above expression of $z = z(\eta)$ yields the expression of $\frac{1}{z}\frac{d^2z}{d\eta^2}$, which is essential for the Mukhanov-Sasaki equation,
\begin{align}
\frac{1}{z}\frac{d^2z}{d\eta^2}=&\frac{\xi(\xi-1)}{\eta^2}
\left[1 + \frac{2\delta(\delta-\rho)\left\{(1-\rho)^2 + 2n(1+\rho-\delta)\right\}}
{\rho(1-\rho)(4n-\rho)(2n+1-\rho)}\left(\frac{R}{R_0}\right)^{\delta-\rho}\right]
\label{sp7}
\end{align}
with $\xi = \frac{(2n+1-\rho)}{(1-2n)}$. Recall $\rho = \frac{1}{4}\left[3 - 2n - \sqrt{1 + 4n(5+n)}\right]$ and 
$\delta = \frac{1}{4}\left[3 - 2n + \sqrt{1 + 4n(5+n)}\right]$, which clearly indicate that $\delta - \rho$ is a positive quantity. Thus the 
term containing $\left(R/R_0\right)^{\delta-\rho}$ within the parenthesis in \ref{sp7} can be safely 
considered to be small during the late contracting era as $R \rightarrow 0$ at $t \rightarrow -\infty$. 
As a result, $\frac{1}{z}\frac{d^2z}{d\eta^2}$ becomes proportional to $1/\eta^2$ i.e., $\frac{1}{z}\frac{d^2z}{d\eta^2} = \sigma/\eta^2$ with,
\begin{align}
\sigma = \xi(\xi-1)\left[1 + \frac{2\delta(\delta-\rho)\left\{(1-\rho)^2 + 2n(1+\rho-\delta)\right\}}
{\rho(1-\rho)(4n-\rho)(2n+1-\rho)}\left(\frac{R}{R_0}\right)^{\delta-\rho}\right]~~.
\label{spnew}
\end{align}
which is approximately a constant in the era, when the primordial perturbation modes generate deep inside the Hubble radius. In 
effect along with the fact that $c_s^2 = 1$, the Mukhanov \ref{sp5} can be solved as follows,
\begin{align}
v(k,\eta) = \frac{\sqrt{\pi|\eta|}}{2} \left[c_1(k)H_{\omega}^{(1)}(k|\eta|) +
c_2(k)H_{\omega}^{(2)}(k|\eta|)\right]\, ,
\label{sp8}
\end{align}
with $\omega = \sqrt{\sigma + \frac{1}{4}}$ and $c_1$ and $c_2$ are integration constants which can be determined from the initial 
Bunch-Davies condition. The consideration of Bunch-Davies vacuum initially, leads to these integration constants as $c_1 = 0$ and $c_2 =1$ 
respectively. Using the solution of $v_k(\eta)$, we immediately evaluate the power spectrum (defined for the Bunch-Davies vacuum state) 
corresponding to the $k$-th scalar perturbation mode, which is defined as follows,
\begin{align}
P_{\Psi}(k,\eta) = \frac{k^3}{2\pi^2}\left|\Psi_k(\eta)\right|^2
= \frac{k^3}{2\pi^2}\left|\frac{\sqrt{\pi|\eta|}}{2z(\eta)}H_{\omega}^{(2)}(k|\eta|)\right|^2
\label{sp9}
\end{align}
The horizon crossing of the mode $k$ is given by $k = \left|aH\right|$ which, due to \ref{late contracting expressions}, take the following form,
\begin{eqnarray}
 k = a_0^n\left|\frac{2n}{t_h^{1-2n}}\right| = \frac{1}{|\eta_h|}\left(\frac{2n}{1-2n}\right)~~,
 \label{hc1}
\end{eqnarray}
where the suffix 'h' denotes the horizon crossing instant and in the second expression, we use the aforementioned relation of $\eta = \eta(t)$. 
\ref{hc1} leads to the horizon crossing time for the large scale modes, in particular for $k = 0.002\mathrm{Mpc}^{-1}$ (around which we will determine 
the observable quantities), as 
\begin{eqnarray}
 \left|\eta_h\right| = \left(\frac{2n}{1-2n}\right)\left(\frac{1}{0.002}\right)\mathrm{Mpc} \approx 10^{17}\mathrm{sec.} \approx -10\mathrm{By}~~.
 \label{hc2}
\end{eqnarray}
Therefore, the large scale modes crosses the horizon at $\sim -10\mathrm{By}$, i.e at deep contracting era. This justifies our consideration 
to use the late contracting expressions of $H(t)$ and $F(R)$ (from \ref{late contracting expressions} and (\ref{FR solution-late contracting})) to evaluate 
the observable quantities. At the horizon crossing of the large scale modes, the Ricci scalar 
acquires $\left|R\right| \sim 10^{-3}\mathrm{By}^{-2}$. 
Moreover, in the present context, the sub-Hubble and super-Hubble scale from \ref{hc1} are given by,
\begin{eqnarray}
 k\left|\eta\right|&>&\frac{2n}{(1-2n)}~~~;~~~~\mathrm{sub~Hubble~scale}\nonumber\\
 k\left|\eta\right|&<&\frac{2n}{(1-2n)}~~~;~~~~\mathrm{super~Hubble~scale}
 \label{sub and super}
\end{eqnarray}
respectively. Here we would like to mention that the factor $2n/(1-2n)$ is less than unity for $n < 1/4$ (see the condition $C1$), 
and thus the superhorizon limit can be equivalently expressed as $k\left|\eta\right| \ll 1$. 
In such superhorizon limit, the scalar power spectrum of \ref{sp9} becomes,
\begin{align}
P_{\Psi}(k,\eta) = \left[\frac{1}{2\pi}\frac{1}{z|\eta|}
\frac{\Gamma(\omega)}{\Gamma(3/2)}\right]^2 \left(\frac{k|\eta|}{2}\right)^{3 - 2\omega}\, .
\label{sp10}
\end{align}
By using Eq.~(\ref{sp10}), we can determine the spectral index of the primordial curvature perturbations (denoted by $n_s$). Before proceeding 
to calculate $n_s$, we will consider first the tensor power spectrum, which is necessary for evaluating the tensor-to-scalar ratio.


\subsection{Tensor perturbation}
In this section we consider the tensor
perturbation on the FRW metric background which is defined as follows,
\begin{eqnarray}
 ds^2 = -dt^2 + a(t)^2\left(\delta_{ij} + h_{ij}\right)dx^idx^j\, ,
 \label{ten per metric}
\end{eqnarray}
where $h_{ij}(t,\vec{x})$ is the tensor perturbation. The variable $h_{ij}(t,\vec{x})$
is itself a gauge invariant quantity, and the tensor perturbed action up to quadratic order is given by \cite{Hwang:2005hb,Noh:2001ia,Hwang:2002fp},
\begin{eqnarray}
 \delta S_{h} = \sum_{L,R}\int dt d^3\vec{x} a(t) z_{L,R}(t)^2\left[\dot{h}_{ij}\dot{h}^{ij}
 - \frac{1}{a^2}\left(\partial_lh_{ij}\right)^2\right]\, ,
 \label{ten per action}
\end{eqnarray}
where the suffix 'L' and 'R' characterize the polarization of the tensor perturbation, in particular the left and right polarization states rspectively. 
The factor $z_{L,R}(t)$, in the Chern-Simons F(R) theory i.e the case of the present context, has the following form \cite{Hwang:2005hb},
\begin{eqnarray}
 z_L^2(t) = \left(\frac{1}{\kappa}\right)a^2(t)F'(R)\left\{1 - \frac{2\dot{\nu}(R)k}{aF'(R)}\right\}~~,\nonumber\\
 z_R^2(t) = \left(\frac{1}{\kappa}\right)a^2(t)F'(R)\left\{1 + \frac{2\dot{\nu}(R)k}{aF'(R)}\right\}~~,
 \label{ten per z}
\end{eqnarray}
with $\nu(R)$ being the CS coupling function (see \ref{action}) and an overdot denotes $\frac{d}{dt}$. It may be observed that the CS term 
has considerable effects on the tensor perturbed action or equivalently on the dynamical evolution of the tensor perturbation variable. In particular, 
due to the presence of $\nu(R)$, the left and right polarization modes of gravity waves evolve differently, 
unlike to the case of vacuum F(R) model where both the tensor polarization get similar evolution. Such difference of the tensor perturbation evolution 
between the CS corrected F(R) and the vacuum F(R) theory reflect on the primordial observable quantity, particularly on the tensor to scalar ratio, as we will 
demonstrate at some stage.

\ref{ten per action} depicts that the speed of the tensor perturbation propagation is $c_T^2 = 1$ for the polarization states. Here we would to mention 
that the unit speed of the gravitational waves is consistent with the event GW170817, according to which the gravitational waves have same speed 
with the electromagnetic waves, i.e unity in natural units. In order to evaluate 
$z_{L,R}(t)$, we consider $\nu(R)$ (having the mass dimension [-2]) to be a power law form of the Ricci scalar, i.e
\begin{eqnarray}
 \nu(R) = \frac{1}{\left|R_0\right|(m+1)}\left(\frac{R}{R_0}\right)^{m+1}~~,
 \label{nu-form}
\end{eqnarray}
with $m$ being a parameter. 
As demonstrated earlier, the large scale modes cross the Hubble horizon during the deep contracting era (in particular, $\sim -10^{17}\mathrm{sec.}$), 
due to which we consider the Hubble parameter and $F(R)$ from \ref{late contracting expressions} and \ref{FR solution-late contracting} respectively. 
In effect and in conjunction with the above form of $\nu(R)$, we determine various terms present in $z_{L,R}(t)$ in \ref{ten per z}:
\begin{eqnarray}
a(t)\sqrt{F'(R)} = \frac{a_0^n\left[12n(1-4n)\right]^n}{\left|R_0\right|^n\left(R/R_0\right)^{n+1/2-\rho/2}}
\left\{1 + \frac{\delta}{\rho}\left(\frac{R}{R_0}\right)^{\delta - \rho}\right\}^{1/2}
\label{term-1}
\end{eqnarray}
and
\begin{eqnarray}
 \frac{2\dot{\nu}(R)k}{aF'(R)} = \frac{8n}{\rho\left[12n(1-4n)\right]}\left(\frac{R}{R_0}\right)^{3-\rho+m}
 \label{term-2}
\end{eqnarray}
respectively. To derive the above expression, the mode-momentum $k$ is evaluated from its horizon crossing condition as,
\begin{eqnarray}
 k = \left|aH\right| = \frac{2na_0^n\left|R_0\right|^{-n+1/2}}{\left[12n(1-4n)\right]^{-n+1/2}}
 \left(\frac{R}{R_0}\right)^{-n+1/2}~~.
 \label{mode-momentum}
\end{eqnarray}
\ref{term-1} and \ref{term-2} lead to the $z_{L,R}$ from \ref{ten per z} as,
\begin{eqnarray}
 z_{L,R} = \left(\frac{a_0^n\left[12n(1-4n)\right]^n}{\left|R_0\right|^n\left(R/R_0\right)^{n+1/2-\rho/2}}\right)
 \left\{1 + \frac{\delta}{2\rho}\left(\frac{R}{R_0}\right)^{\delta - \rho} 
 \mp \frac{4n}{\rho\left[12n(1-4n)\right]}\left(\frac{R}{R_0}\right)^{3-\rho+m}\right\}~~,
 \label{z-final-1}
\end{eqnarray}
where we consider the leading order terms of $\left(R/R_0\right)$ as $\frac{R}{R_0} < 1$ during the horizon crossing of the primordial 
perturbation modes. For convenience, we parametrize $m = \rho - 3 + (\delta-\rho)(1-g)$ in respect to a new parameter $g$. With such parametrization, 
\ref{z-final-1} becomes,
\begin{eqnarray}
 z_{L,R} = \left(\frac{a_0^n\left[12n(1-4n)\right]^n}{\left|R_0\right|^n\left(R/R_0\right)^{n+1/2-\rho/2}}\right)
 \left\{1 + \frac{\delta}{2\rho}\left(\frac{R}{R_0}\right)^{\delta - \rho} 
 \mp \frac{4n}{\rho\left[12n(1-4n)\right]}\left(\frac{R}{R_0}\right)^{(\delta-\rho)(1-g)}\right\}~~.
 \label{z-final-2}
\end{eqnarray}
The parameter $g$ reflects the possible effects of the CS coupling function in the above expression of $z_{L,R}$. In the vacuum F(R) theory, 
$z_{(F)} = \left(\frac{a_0^n\left[12n(1-4n)\right]^n}{\left|R_0\right|^n\left(R/R_0\right)^{n+1/2-\rho/2}}\right)
 \left\{1 + \frac{\delta}{2\rho}\left(\frac{R}{R_0}\right)^{\delta - \rho}\right\}$ which, by comparing with 
 \ref{z-final-2}, clearly indicates that the difference between the CS corrected F(R) and the vacuum F(R) theory is controlled by $g$. Thereby 
 it seems that the parameter $g$ plays a crucial role in the present context and thus we need to scan it carefully before proceeding further. 
 Depending on various values of $g$, below we give a list of $z_{L,R}$ in the leading order of $R/R_0$,
 \begin{eqnarray}
  z_{L,R}&=&\left(\frac{a_0^n\left[12n(1-4n)\right]^n}{\left|R_0\right|^n\left(R/R_0\right)^{n+1/2-\rho/2}}\right)
 \left\{1 + \frac{\delta}{2\rho}\left(\frac{R}{R_0}\right)^{\delta - \rho}\right\}~~~~~~;~~~~~\mathrm{for}~g < 0\label{1}\\
 z_{L,R}&=&\left(\frac{a_0^n\left[12n(1-4n)\right]^n}{\left|R_0\right|^n\left(R/R_0\right)^{n+1/2-\rho/2}}\right)
 \left\{1 + \left(\frac{\delta}{2\rho} \mp \frac{4n}{\rho\left[12n(1-4n)\right]}\right)\left(\frac{R}{R_0}\right)^{\delta - \rho}\right\}
 ~~~;~~~\mathrm{for}~g = 0\label{2}\\
 z_{L,R}&=&\left(\frac{a_0^n\left[12n(1-4n)\right]^n}{\left|R_0\right|^n\left(R/R_0\right)^{n+1/2-\rho/2}}\right)
 \left\{1 \mp \frac{4n}{\rho\left[12n(1-4n)\right]}\left(\frac{R}{R_0}\right)^{(\delta-\rho)(1-g)}\right\}~~;~~\mathrm{for}~0 < g < 1\label{3}
 \end{eqnarray}
 Now we need to investigate qualitatively that which of the above expressions of $z_{L,R}$ will be appropriate for evaluating 
 the observable quantities in the present context. For this purpose, we may recall that the vacuum F(R) theory is not consistent with the Planck results 
 of primordial observable quantities in the background of a non-singular bounce where $a(t) \sim t^{2n}$ during the early contracting era 
 \cite{Odintsov:2014gea,Nojiri:2019lqw}. 
 In particular, the scalar and tensor perturbation amplitudes in vacuum F(R) bounce scenario become comparable 
 to each other and thus the tensor-to-scalar ratio comes as order of unity, which is not compatible with the Planck constraint. Thereby 
 to get a viable bounce scenario, either the tensor perturbation amplitude needs to be suppressed or the scalar perturbation amplitude needs to be 
 enhanced in comparison to the vacuum F(R) theory so that 
 the tensor-to-scalar ratio becomes less than unity and comes within the Planck constraint. In 
 this regard, comparing the above three expressions of $z_{L,R}$ with $z_{(F)}$, we may argue that 
 in the present context of CS corrected F(R) theory, the $z_{L,R}$ and consequently the tensor perturbation evolution get 
 considerably different than the vacuum F(R) case, when the parameter $g$ lies within $0 < g < 1$ (i.e \ref{3}); thus for $0<g<1$, there is a 
 possibility to get viable observable quantities in the CS corrected F(R) model. Based on these arguments, we consider 
 $0 < g < 1$, in which case the $z_{L,R}$ is given by \ref{3}. 
 
\begin{figure}[!h]
\begin{center}
 \centering
 \includegraphics[scale=0.70]{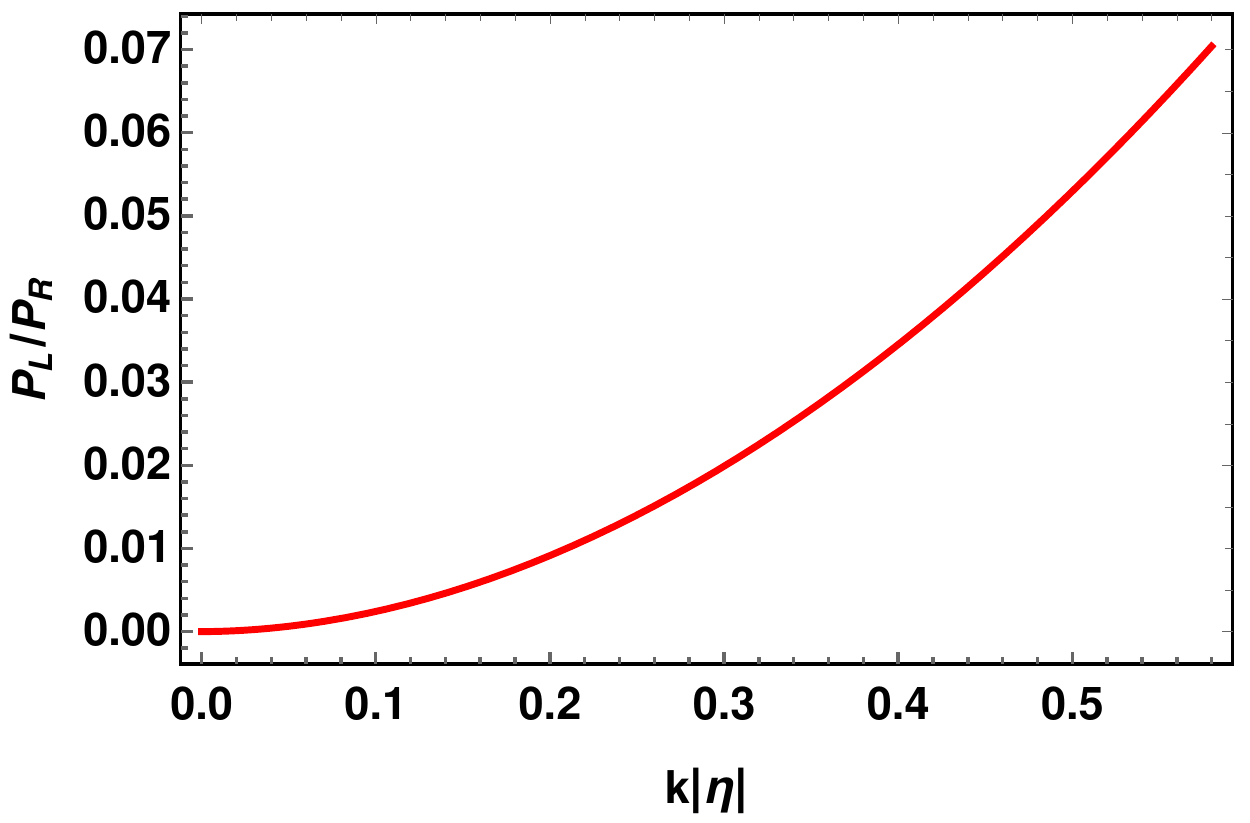}
 \caption{$P_L/P_R$ vs. $k\left|\eta\right|$ in the superhorizon scale, i.e during $k\left|\eta\right| < \frac{2n}{1-2n}$. 
 The plot corresponds to $n = 0.185$ and $g = 0.5$.}
 \label{plot-mode-comparison}
\end{center}
\end{figure}

Similar to scalar perturbation, the Mukhanov-Sasaki variable for 
tensor perturbation is defined as $v_{\lambda} = z_{\lambda}~h_{\lambda}$ (with $\lambda = L,R$) which, 
upon performing the Fourier transformation, satisfies the following equation,
\begin{align}
\frac{d^2v_{\lambda}(k,\eta)}{d\eta^2}
+ \left(k^2 - \frac{1}{z_{\lambda}(\eta)}\frac{d^2z_{\lambda}}{d\eta^2}\right)v_{\lambda}(k,\eta) = 0~~.
\label{tp4}
\end{align}
By using Eq.~(\ref{sp6}), we evaluate $z_{\lambda}(\eta)$ and $\frac{1}{z_{\lambda}(\eta)}\frac{d^2z_{\lambda}}{d\eta^2}$ and these read,
\begin{align}
z_{L,R} \propto \left\{1 \mp \frac{4n}{\rho\left[12n(1-4n)\right]}\left(\frac{R}{R_0}\right)^{(\delta-\rho)(1-g)}\right\}\times\eta^{(2n+1-\rho)/(1-2n)}
\label{tpnew}
\end{align}
and
\begin{align}
\frac{1}{z_{L,R}}\frac{d^2z_{L,R}}{d\eta^2} = \frac{\xi(\xi-1)}{\eta^2}
\left\{1 \mp \frac{16n(\delta-\rho)(1-g)}{\rho(4n-\rho)\left[12n(1-4n)\right]}\left(\frac{R}{R_0}\right)^{(\delta-\rho)(1-g)}\right\}
\label{tp6}
\end{align}
respectively. Due to the fact that $\delta-\rho$ is positive, the variation of the term in the parenthesis in Eq.~(\ref{tp6}), 
can be regarded to be small in the low-curvature regime or equivalently during the early contracting stage and thus 
$\frac{1}{z_{\lambda}}\frac{d^2z_{\lambda}}{d\eta^2}$ becomes proportional to 
$1/\eta^2$ that is $\frac{1}{z_{\lambda}}\frac{d^2z_{\lambda}}{d\eta^2} = \sigma_{\lambda}/\eta^2$ (with $\lambda = L,R$), where
\begin{align}
\sigma_{L,R} = \xi(\xi-1)
\left\{1 \mp \frac{16n(\delta-\rho)(1-g)}{\rho(4n-\rho)\left[12n(1-4n)\right]}\left(\frac{R}{R_0}\right)^{(\delta-\rho)(1-g)}\right\}~~,
\label{tp9}
\end{align}
and recall $\xi = \frac{(2n+1-\rho)}{(1-2n)}$. The above expressions yield the tensor power spectrum, defined with the initial Bunch-Davies vacuum state 
at the deep sub-Hubble radius, so we have,
\begin{eqnarray}
 P_{h}(k,\eta) = P_{L}(k,\eta) + P_{R}(k,\eta)
 \label{tp10}
\end{eqnarray}
with
\begin{eqnarray}
P_{L}(k,\eta) = \left[\frac{1}{2\pi}\frac{1}{z_L|\eta|}\frac{\Gamma(\Omega_L)}{\Gamma(3/2)}\right]^2 \left(\frac{k|\eta|}{2}\right)^{3 - 2\Omega_L}~~,\nonumber\\
P_{R}(k,\eta) = \left[\frac{1}{2\pi}\frac{1}{z_R|\eta|}\frac{\Gamma(\Omega_R)}{\Gamma(3/2)}\right]^2 \left(\frac{k|\eta|}{2}\right)^{3 - 2\Omega_R}~~.
\label{tp11}
\end{eqnarray}
The factor $\Omega_{L,R} = \sqrt{\sigma_{L,R} + \frac{1}{4}}$ where $\sigma_{L,R}$ is defined in Eq.~(\ref{tp9}). It may be 
observed that the left and right polarization modes of the tensor perturbation have different power spectra at a given 
$(k,\eta)$, due to the fact that $\Omega_L \neq \Omega_R$ which actually inherits from the CS coupling function. In particular, 
\ref{tp9} indicates $\Omega_R > \Omega_L$, which in turn makes $P_L(k,\eta)$ suppressed compared to the $P_R(k,\eta)$ in the 
superhorizon scale. We give the plot of $P_L/P_R$ vs. $k\left|\eta\right|$ in the superhorizon limit, i.e during $k\left|\eta\right| < \frac{2n}{1-2n}$, 
in \ref{plot-mode-comparison} which corresponds to $n = 0.185$ and $g = 0.5$ (such values of $n$ and $g$ are compatible in respect to the 
Planck results of $n_s$ and $r$, as we will demonstrate soon). \ref{plot-mode-comparison} clearly depicts that the $P_L(k,\eta)$ is indeed 
suppressed than the $P_R(k,\eta)$. 

Now we can explicitly confront the model at hand with the latest Planck observational data \cite{Akrami:2018odb}, so we 
calculate the spectral index of the primordial curvature perturbations $n_s$ and the tensor-to-scalar ratio $r$, which are defined as follows,
\begin{align}
n_s = 1 + \left. \frac{\partial \ln{P_{\Psi}}}{\partial
\ln{k}}\right|_{h.c} \, , \quad
r = \left. \frac{P_{h}(k,\eta)}{P_{\Psi}(k,\eta)}\right|_{h.c}~~,
\label{obs1}
\end{align}
where $P_{\Psi}(k,\eta)$ and $P_h(k,\eta)$ are obtained in \ref{sp10} and \ref{tp10} respectively, and the suffix 'h.c' denotes the horizon crossing 
instant when the mode $k$ satisfies $k = \left|aH\right|$. From \ref{sp10}, the scalar spectral 
index comes with the following expression,
\begin{align}
n_s = 4 - \sqrt{1 + 4\sigma}
\label{obs2}
\end{align}
with $\sigma$ being given in \ref{spnew}.

\begin{figure}
\begin{center}
 \centering
 \includegraphics[scale=0.55]{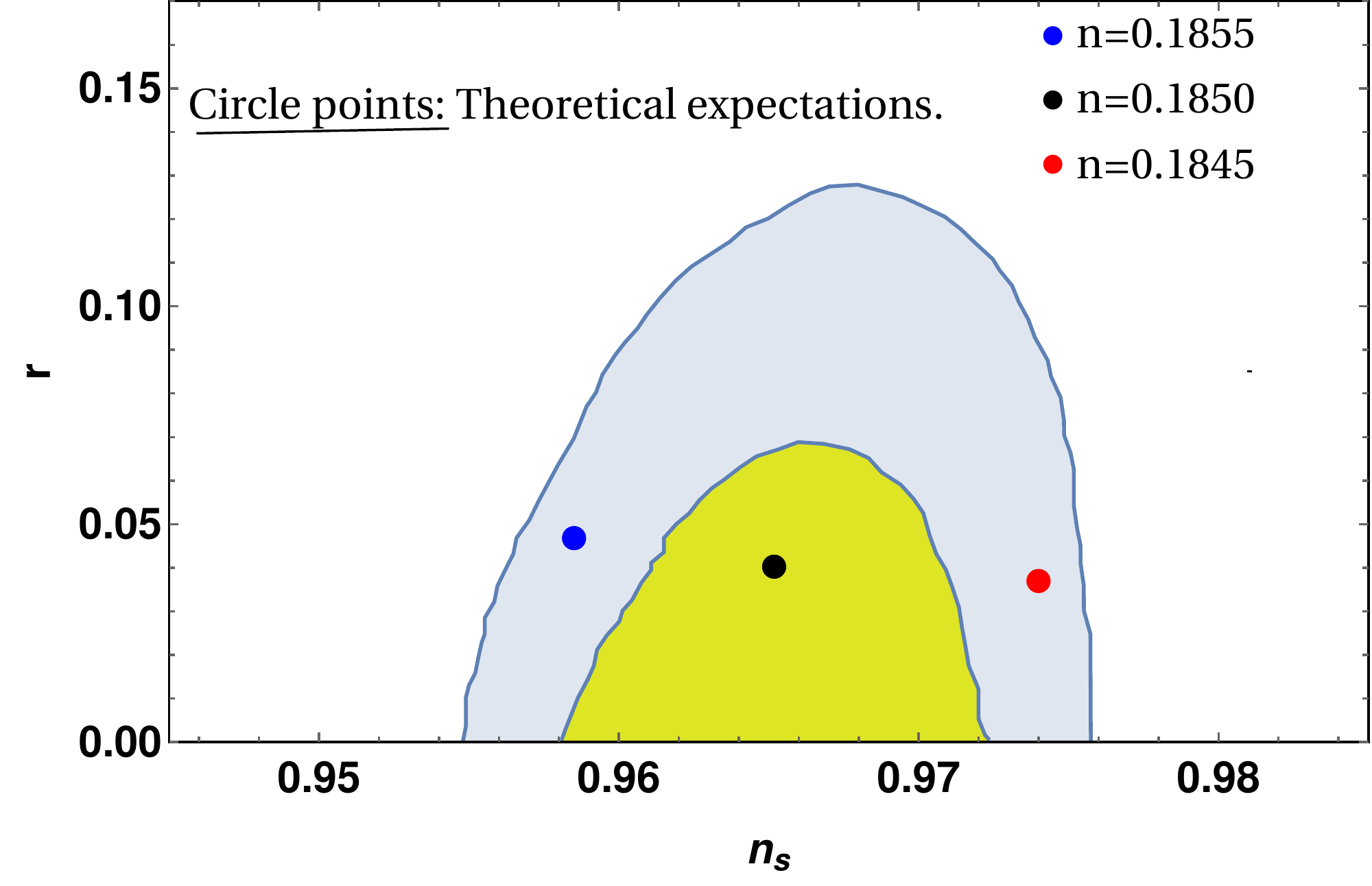}
 \caption{$1\sigma$ (yellow) and $2\sigma$ (light blue) contours for Planck 
 2018 results \cite{Akrami:2018odb}, on $n_s-r$ plane. 
 Additionally, we present the predictions of the present bounce scenario with 
 $n = 0.1855$ (blue point), $n = 0.185$ (black point) and $n = 0.1845$ (red point).}
 \label{plot-observable}
\end{center}
\end{figure}

It may be noticed that $n_s$ depends on $\frac{R_h}{R_0}$ and $n$, while $r$ depends on 
$\frac{R_h}{R_0}$, $n$ and $g$. The dependency of $r$ on the parameter $g$ comes from the fact that the CS coupling function, which 
contains the parameter $g$, affects the tensor perturbation only. The $R_h$ denotes the Ricci scalar at the horizon crossing 
instant of the large scale modes (in particular $k = 0.002\mathrm{Mpc}^{-1}$) on which we are interested to evaluate the observable quantities. As mentioned 
earlier, the mode $k = 0.002\mathrm{Mpc}^{-1}$ crosses the horizon at $t_h \approx -10\mathrm{By}$ and thereby the corresponding 
Ricci scalar is given by $R(t_h) = \frac{-12n(1-4n)}{t_h^2} \approx -0.12n(1-4n)$. Taking $R_0 = -1\mathrm{By}^{-2}$ (recall $R_0$ is negative, 
see the discussion after \ref{F-prime-R}), we get $\frac{R_h}{R_0} = 0.12n(1-4n)$. Thus, as a whole, the spectral index 
depends only on $n$ and the tensor-to-scalar ratio depends on $n$ and $g$. 
With this information, we now directly confront the theoretical expressions of scalar spectral index \ref{obs2}
and tensor-to-scalar ratio \ref{obs1} derived from the present model with the Planck 2018 constraints 
\cite{Akrami:2018odb}. In particular,  we estimate the allowed values of $n$ and $g$ 
which in turn can give rise to $n_s$ and $r$ in agreement with the Planck data. 
This is presented in \ref{plot-observable} where we compute $n_s$ and $r$ for three choices 
of $n$, viz. $n = 0.1855$ (blue point), $n = 0.185$ (black point) and $n = 0.1845$ (red point) 
with $g = 0.5$. The allowed values of $n_s$ and $r$ from 
Planck data within $1-\sigma$ and $2-\sigma$ constraints are illustrated by the yellow and the blue regions 
respectively in \ref{plot-observable}. We note that with $g = 0.5$ and all the three aforesaid values 
of $n$, the model estimated $n_s$ and $r$ are within the $1-\sigma$ or $2-\sigma$ constraints reported by Planck 2018 data. 
Thereby in the present context, the scalar spectral index and the tensor-to-scalar ratio are simultaneously compatible with the latest Planck 2018 
constraints. On contrary, here we would like to mention that in the vacuum F(R) model, the observable quantities like $n_s$ and $r$ are not 
simultaneously compatible with the Planck results in the background of a non-singular bounce where $a(t) \sim t^{2n}$ during early contracting stage. 
In particular, the scalar and tensor perturbation amplitudes in the vacuum F(R) bounce model become comparable to each other and thus the tensor-to-scalar 
ratio comes as order of unity which is excluded from the Planck data. However, in the Chern-Simons corrected F(R) theory, the CS coupling function 
considerably affects the tensor perturbation evolution, keeping intact the scalar type perturbation with that of in the vacuum F(R) case. In effect, 
the tensor perturbation amplitude in the Chern-Simons generalized F(R) bounce model gets suppressed compared to the vacuum F(R) case, 
and as a result, the tensor-to-scalar ratio in the present context becomes less than unity and comes within the Planck constraints. 
Such effects of CS coupling function on the tensor type perturbation is in agreement 
with \cite{Odintsov:2019mlf} where some of our authors showed the similar effects of the CS term in the context of an inflationary background spacetime. 
It has been showed in \cite{Odintsov:2019mlf}, that a simple power-law F(R) inflationary model, 
in particular $F(R) = R + \alpha R^{n}$ (with $n \approx 1.817$), 
without the CS term yields the correct value of the scalar spectral index, but the tensor-to-scalar ratio comes as 
$r \approx 0.24$ which is excluded from the Planck data \cite{Akrami:2018odb}; however the 
inclusion of CS term in this F(R) model reduces the value of the tensor-to-scalar ratio without affecting 
the scalar spectral index, and thus the inflationary parameters of the said $R+\alpha R^{n}+\mathrm{CS}$ model become compatible with the observations.\\ 

Before concluding, here we would like to mention that in regard to the observable parameters related to the early epoch of the universe, 
we have evaluated the scalar spectral index ($n_s$) and tensor-to-scalar ratio ($r$) of the 
primordial perturbations, which appears to be consistent with the Planck data, well within the 1-$\sigma$ and 2-$\sigma$ region, 
for $0.1845 \lesssim n \lesssim 0.1855$ 
(recall, $n$ appears in the power exponent of the scale factor). Therefore, in future, if $n_s$ and $r$ are further improved by the Planck collaborations, 
the parameter $n$ and hence the bouncing behaviour of the scale factor can be better constrained. 
Beside the scalar spectral index and the tensor-to-scalar ratio (that are 
related to the two point correlators of scalar and tensor perturbations respectively), the bounce 
scenario can also be examined from the corresponding higher point correlators of primordial perturbations, in particular, by 
estimating the theoretical expectations of various non-linear parameters 
(like $f_\mathrm{NL}$, $\tau_\mathrm{NL}$ etc, which represent the amplitudes of the bispectrum 
and trispectrum respectively \cite{Akrami:2018odb,Cai:2009fn,Agullo:2020cvg}) with the respective Planck data. Furthermore, it is also important to 
study the signatures of lower scale modes of tensor perturbation, and investigate them with respect to the sensitivity of 
various gravitational waves (GWs) observatories. The proposed GWs observatories may include advanced LIGO ($10-10^3$ Hz), ET ($1-10^4$ Hz), 
BBO ($10^{-3}-10$ Hz), DECIGO ($10^{-3}-1$ Hz), eLISA ($10^{-5}-1$ Hz), and SKA ($10^{-9}-10^{-6}$ Hz) \cite{LIGOScientific:2016jlg,
Janssen:2014dka,Amaro-Seoane:2012aqc}. The presence of the parity violating Chern-Simons term in the model distinguishes the evolution of the two polarization modes 
of tensor perturbation and leads to the generation of chiral gravitational waves, which may have non-trivial imprints on today's GWs spectrum. 
Therefore the evolution of primordial gravitational waves in the background of the bounce scenario discussed here may be compared with the future GWs spectrum from various observatories to provide a unique way to constraint our model. We hope to 
address these issues in our future work.

\section{Conclusion}
In this work, we proposed an unified cosmological scenario of a non-singular bounce to a dark energy (DE) epoch in the context of 
Chern-Simons corrected F(R) gravity theory, where the Chern-Simons coupling function is assumed to have a power law behaviour with the 
Ricci scalar. Using the reconstruction technique, we analytically determine the form of F(R) during the late contracting era. Using such analytic solution, 
and in addition, by employing suitable boundary conditions, we numerically solve the gravitational equation and evaluate the F(R) for 
the entire possible range of the cosmic time, which clearly depicts that the F(R) matches with the Einstein gravity in the low curvature 
regime, while it deviates from the usual Einstein gravity as the scalar curvature acquires larger and larger values. 
The form of F(R) leads to an unified cosmological 
scenario of a non-singular bounce to a dark energy epoch, in particular, from a bounce to a deceleration stage having a 
matter-like evolution during some regime of the deceleration stage and from the deceleration phase to a late time acceleration era. The effective 
EoS of the dark energy epoch acquires the value $\omega_\mathrm{eff} = -0.997$ at present time, which is indeed compatible with 
the results provided by Planck+SNe+BAO data. Moreover the 
model predicts a finite time future singularity of the universe around $30\mathrm{By}$ when the scale factor, the effective energy density and 
the effective pressure are found to diverge, and thus the singularity is a Type-I type of singularity. However, since the present age of our universe is 
$t_p \approx 13.5\mathrm{By}$, i.e the Type-I singularity occurs at far future from the present age, 
we may argue that the current model satisfactorily describes a singular free evolution of the universe up-to the cosmic time 
$t \gtrsim t_p$. Here it deserves mentioning that the bounce in the present context is an asymmetric bounce, in particular, the comoving Hubble radius 
monotonically increases with cosmic time and asymptotically diverges at distant past, while 
it decreases with time at the present epoch of the universe. Due to such evolution of the Hubble horizon, the primordial perturbation modes 
generate at distant past far away from the bounce when all the relevant perturbation modes lie within the horizon. Correspondingly the scalar and tensor 
perturbations power spectra are determined, which in turn leads to the primordial observable quantities like the spectral index of the scalar 
curvature perturbation ($n_s$) and the tensor-to-scalar ratio ($r$). The theoretical expectations of $n_s$ and $r$ in the present context 
are found to be simultaneously compatible with the latest Planck 2018 constraints. In this regard, the Chern-Simons term proves to play 
an important role in making the observable quantities, particularly the tensor-to-scalar ratio, consistent with the Planck data. Actually in the case of 
vacuum F(R) model in the background of a non-singular bounce, the scalar and tensor perturbation amplitudes are comparable to each other and thus 
the tensor-to-scalar ratio becomes order of unity which is indeed excluded from the Planck results. However, in the F(R) model generalized by 
the Chern-Simons term, the perturbations evolution get considerably affected due to the presence of the CS term, 
in particular the tensor perturbation amplitude gets suppressed and the scalar perturbation remains intact compared 
to that of in the vacuum F(R) case. As a result, the tensor-to-scalar ratio in the Chern-Simons corrected F(R) bounce model becomes less than unity 
and moreover it comes within the Planck constraints for a suitable parametric regime.

In summary, the present model provides an unified cosmological scenario of a non-singular bounce to viable dark energy epoch in the Chern-Simons 
generalized F(R) theory, where the Chern-Simons term plays a crucial role in regard to the compatibility of the primordial 
observable quantities with the Planck results.

\subsection*{Acknowledgments}
This work was supported in part by MINECO (Spain), project PID2019-104397GB-I00 (SDO). 
TP sincerely acknowledges the hospitality by ICE-CSIC/IEEC (Barcelona, Spain), where this work was generated during his visit. 
The work was  supported by the Ministry of Education and Science of the Republic of Kazakhstan, Grant AP09261147 (RM). 



\begin{thebibliography}{99}
 \bibitem{guth}
A.H. Guth;  Phys.Rev. D23 347-356 (1981).

\bibitem{Linde:2005ht}
  A.~D.~Linde,
  Contemp.\ Concepts Phys.\  {\bf 5} (1990) 1
  [hep-th/0503203].
  
  
  





\bibitem{Langlois:2004de}
  D.~Langlois,
  hep-th/0405053.


\bibitem{Riotto:2002yw}
  A.~Riotto,
  ICTP Lect.\ Notes Ser.\  {\bf 14} (2003) 317
  [hep-ph/0210162].


\bibitem{barrow1}
J. D. Barrow and P. Saich; Class. Quantum Grav. 10, 279 (1993).

\bibitem{barrow2}
J. D. Barrow and J. P. Mimoso; Phys. Rev. D 50, 3746 (1994).



\bibitem{Baumann:2009ds}
D.~Baumann,
doi:10.1142/9789814327183 0010
[arXiv:0907.5424 [hep-th]].


\bibitem{Brandenberger:2012zb}
R.~H.~Brandenberger,
arXiv:1206.4196 [astro-ph.CO].




\bibitem{Brandenberger:2016vhg}
R.~Brandenberger and P.~Peter,
arXiv:1603.05834 [hep-th].


\bibitem{Battefeld:2014uga}
 D.~Battefeld and P.~Peter,
 Phys.\ Rept.\ {\bf 571} (2015) 1
 doi:10.1016/j.physrep.2014.12.004
 [arXiv:1406.2790 [astro-ph.CO]].


\bibitem{Novello:2008ra}
M.~Novello and S.~E.~P.~Bergliaffa,
 ``Bouncing Cosmologies,''
Phys.\ Rept.\ {\bf 463} (2008) 127
doi:10.1016/j.physrep.2008.04.006
[arXiv:0802.1634 [astro-ph]].


\bibitem{Cai:2014bea}
Y.~F.~Cai,
Sci.\ China Phys.\ Mech.\ Astron.\  {\bf 57} (2014) 1414
doi:10.1007/s11433-014-5512-3
[arXiv:1405.1369 [hep-th]].


\bibitem{Cai:2016thi}
Y.~Cai, Y.~Wan, H.~G.~Li, T.~Qiu and Y.~S.~Piao,
JHEP \textbf{01} (2017), 090
doi:10.1007/JHEP01(2017)090
[arXiv:1610.03400 [gr-qc]].




\bibitem{deHaro:2015wda}
J.~de Haro and Y.~F.~Cai,
Gen.\ Rel.\ Grav.\ {\bf 47} (2015) no.8, 95
doi:10.1007/s10714-015-1936-y
[arXiv:1502.03230 [gr-qc]].






\bibitem{Lehners:2011kr}
J.~L.~Lehners,
Class.\ Quant.\ Grav.\ {\bf 28} (2011) 204004
doi:10.1088/0264-9381/28/20/204004
[arXiv:1106.0172 [hep-th]].



\bibitem{Lehners:2008vx}
J.~L.~Lehners,
Phys.\ Rept.\ {\bf 465} (2008) 223
doi:10.1016/j.physrep.2008.06.001
[arXiv:0806.1245 [astro-ph]].


\bibitem{Cai:2016hea}
Y.~F.~Cai, A.~Marciano, D.~G.~Wang and E.~Wilson-Ewing,
Universe {\bf 3} (2016) no.1,  1 doi:10.3390/Universe3010001
[arXiv:1610.00938 [astro-ph.CO]].


\bibitem{Colin:2017dwv}
S.~Colin and N.~Pinto-Neto,
Phys. Rev. D \textbf{96} (2017) no.6, 063502
doi:10.1103/PhysRevD.96.063502
[arXiv:1706.03037 [gr-qc]].





\bibitem{Cattoen:2005dx}
C.~Cattoen and M.~Visser,
Class.\ Quant.\ Grav.\ {\bf 22} (2005) 4913
doi:10.1088/0264-9381/22/23/001
[gr-qc/0508045].




\bibitem{Li:2014era}
C.~Li, R.~H.~Brandenberger and Y.~K.~E.~Cheung,
Phys.\ Rev.\ D {\bf 90} (2014) no.12, 123535
doi:10.1103/PhysRevD.90.123535
[arXiv:1403.5625 [gr-qc]].




\bibitem{Brizuela:2009nk}
D.~Brizuela, G.~A.~D.~Mena Marugan and T.~Pawlowski,
Class.\ Quant.\ Grav.\ {\bf 27} (2010) 052001
doi:10.1088/0264-9381/27/5/052001
[arXiv:0902.0697 [gr-qc]].



\bibitem{Cai:2013kja}
Y.~F.~Cai, E.~McDonough, F.~Duplessis and R.~H.~Brandenberger,
JCAP {\bf 1310} (2013) 024
doi:10.1088/1475-7516/2013/10/024
[arXiv:1305.5259 [hep-th]].


\bibitem{Quintin:2014oea}
J.~Quintin, Y.~F.~Cai and R.~H.~Brandenberger,
Phys.\ Rev.\ D {\bf 90} (2014) no.6, 063507
doi:10.1103/PhysRevD.90.063507
[arXiv:1406.6049 [gr-qc]].


\bibitem{Cai:2013vm}
Y.~F.~Cai, R.~Brandenberger and P.~Peter,
Class.\ Quant.\ Grav.\ {\bf 30} (2013) 075019
doi:10.1088/0264-9381/30/7/075019
[arXiv:1301.4703 [gr-qc]].


\bibitem{pinto}
Pinto-Neto, N. Bouncing Quantum Cosmology. Universe 2021, 7, 110. https://doi.org/10.3390/universe7040110




\bibitem{Koehn:2015vvy}
M.~Koehn, J.~L.~Lehners and B.~Ovrut,
Phys.\ Rev.\ D {\bf 93} (2016) no.10, 103501
doi:10.1103/PhysRevD.93.103501
[arXiv:1512.03807 [hep-th]].




\bibitem{Nojiri:2016ygo}
S.~Nojiri, S.~D.~Odintsov and V.~K.~Oikonomou,
Phys.\ Rev.\ D {\bf 93} (2016) no.8, 084050
doi:10.1103/PhysRevD.93.084050
[arXiv:1601.04112 [gr-qc]].




\bibitem{Odintsov:2020zct}
S.~D.~Odintsov, V.~K.~Oikonomou and T.~Paul,
Class. Quant. Grav. \textbf{37} (2020) no.23, 235005
doi:10.1088/1361-6382/abbc47
[arXiv:2009.09947 [gr-qc]].





\bibitem{Koehn:2013upa}
M.~Koehn, J.~L.~Lehners and B.~A.~Ovrut,
Phys.\ Rev.\ D {\bf 90} (2014) no.2, 025005
doi:10.1103/PhysRevD.90.025005
[arXiv:1310.7577 [hep-th]].



\bibitem{Battarra:2014kga}
L.~Battarra and J.~L.~Lehners,
JCAP {\bf 1412} (2014) no.12, 023
doi:10.1088/1475-7516/2014/12/023
[arXiv:1407.4814 [hep-th]].


\bibitem{Martin:2001ue}
J.~Martin, P.~Peter, N.~Pinto Neto and D.~J.~Schwarz,
Phys.\ Rev.\ D {\bf 65} (2002) 123513
doi:10.1103/PhysRevD.65.123513
[hep-th/0112128].


\bibitem{Khoury:2001wf}
J.~Khoury, B.~A.~Ovrut, P.~J.~Steinhardt and N.~Turok,
Phys.\ Rev.\ D {\bf 64} (2001) 123522
doi:10.1103/PhysRevD.64.123522
[hep-th/0103239].










\bibitem{Hackworth:2004xb}
J.~C.~Hackworth and E.~J.~Weinberg,
Phys.\ Rev.\ D {\bf 71} (2005) 044014
doi:10.1103/PhysRevD.71.044014
[hep-th/0410142].




\bibitem{Johnson:2011aa}
M.~C.~Johnson and J.~L.~Lehners,
Phys.\ Rev.\ D {\bf 85} (2012) 103509
doi:10.1103/PhysRevD.85.103509
[arXiv:1112.3360 [hep-th]].



\bibitem{Peter:2002cn}
P.~Peter and N.~Pinto-Neto,
Phys.\ Rev.\ D {\bf 66} (2002) 063509
doi:10.1103/PhysRevD.66.063509
[hep-th/0203013].


\bibitem{Gasperini:2003pb}
M.~Gasperini, M.~Giovannini and G.~Veneziano,
Phys.\ Lett.\ B {\bf 569} (2003) 113
doi:10.1016/j.physletb.2003.07.028
[hep-th/0306113].



\bibitem{Creminelli:2004jg}
P.~Creminelli, A.~Nicolis and M.~Zaldarriaga,
Phys.\ Rev.\ D {\bf 71} (2005) 063505
doi:10.1103/PhysRevD.71.063505
[hep-th/0411270].


\bibitem{Lehners:2015mra}
J.~L.~Lehners and E.~Wilson-Ewing,
JCAP {\bf 1510} (2015) no.10, 038
doi:10.1088/1475-7516/2015/10/038
[arXiv:1507.08112 [astro-ph.CO]].






\bibitem{Lehners:2013cka}
J.~L.~Lehners and P.~J.~Steinhardt,
Phys.\ Rev.\ D {\bf 87} (2013) no.12, 123533
doi:10.1103/PhysRevD.87.123533
[arXiv:1304.3122 [astro-ph.CO]].




\bibitem{Cai:2014xxa}
Y.~F.~Cai, J.~Quintin, E.~N.~Saridakis and E.~Wilson-Ewing,
JCAP {\bf 1407} (2014) 033
doi:10.1088/1475-7516/2014/07/033
[arXiv:1404.4364 [astro-ph.CO]].




\bibitem{Cai:2007qw}
Y.~F.~Cai, T.~Qiu, Y.~S.~Piao, M.~Li and X.~Zhang,
JHEP {\bf 0710} (2007) 071
doi:10.1088/1126-6708/2007/10/071
[arXiv:0704.1090 [gr-qc]].







\bibitem{Barrow:2004ad}
J.~D.~Barrow, D.~Kimberly and J.~Magueijo,
Class.\ Quant.\ Grav.\ {\bf 21} (2004) 4289
doi:10.1088/0264-9381/21/18/001
[astro-ph/0406369].

\bibitem{Haro:2015zda}
J.~Haro and E.~Elizalde,
JCAP {\bf 1510} (2015) no.10, 028
doi:10.1088/1475-7516/2015/10/028
[arXiv:1505.07948 [gr-qc]].




\bibitem{Das:2017jrl}
A.~Das, D.~Maity, T.~Paul and S.~SenGupta,
Eur.\ Phys.\ J.\ C {\bf 77} (2017) no.12,  813
doi:10.1140/epjc/s10052-017-5396-2
[arXiv:1706.00950 [hep-th]].




\bibitem{deHaro:2012xj}
J.~de Haro,
JCAP {\bf 1211} (2012) 037
[arXiv:1207.3621 [gr-qc]].










\bibitem{WilsonEwing:2012pu}
E.~Wilson-Ewing,
JCAP {\bf 1303} (2013) 026
doi:10.1088/1475-7516/2013/03/026
[arXiv:1211.6269 [gr-qc]].

\bibitem{Elizalde:2019tee}
E.~Elizalde, S.~D.~Odintsov and T.~Paul,
Eur. Phys. J. C \textbf{80} (2020) no.1, 10
doi:10.1140/epjc/s10052-019-7544-3
[arXiv:1912.05138 [gr-qc]].





\bibitem{Cai:2008qw}
  Y.~F.~Cai, T.~t.~Qiu, R.~Brandenberger and X.~m.~Zhang,
  Phys.\ Rev.\ D {\bf 80} (2009) 023511
  doi:10.1103/PhysRevD.80.023511
  [arXiv:0810.4677 [hep-th]].




\bibitem{Finelli:2001sr}
F.~Finelli and R.~Brandenberger,
Phys.\ Rev.\ D {\bf 65} (2002) 103522
doi:10.1103/PhysRevD.65.103522
[hep-th/0112249].



\bibitem{Cai:2011ci}
Y.~F.~Cai, R.~Brandenberger and X.~Zhang,
Phys.\ Lett.\ B {\bf 703} (2011) 25
doi:10.1016/j.physletb.2011.07.074
[arXiv:1105.4286 [hep-th]].


\bibitem{Haro:2015zta}
J.~Haro and J.~Amor\' os,
PoS FFP {\bf 14} (2016) 163
doi:10.22323/1.224.0163
[arXiv:1501.06270 [gr-qc]].

\bibitem{Cai:2011zx}
Y.~F.~Cai, R.~Brandenberger and X.~Zhang,
JCAP {\bf 1103} (2011) 003
doi:10.1088/1475-7516/2011/03/003
[arXiv:1101.0822 [hep-th]].









\bibitem{Brandenberger:2009yt}
R.~Brandenberger,
Phys.\ Rev.\ D {\bf 80} (2009) 043516
doi:10.1103/PhysRevD.80.043516
[arXiv:0904.2835 [hep-th]].

\bibitem{deHaro:2014kxa}
J.~de Haro and J.~Amoros,
JCAP {\bf 1408} (2014) 025
doi:10.1088/1475-7516/2014/08/025
[arXiv:1403.6396 [gr-qc]].


\bibitem{Odintsov:2014gea}
S.~D.~Odintsov and V.~K.~Oikonomou,
Phys.\ Rev.\ D {\bf 90} (2014) no.12, 124083
doi:10.1103/PhysRevD.90.124083
[arXiv:1410.8183 [gr-qc]].



\bibitem{Qiu:2010ch}
T.~Qiu and K.~C.~Yang,
JCAP {\bf 1011} (2010) 012
doi:10.1088/1475-7516/2010/11/012
[arXiv:1007.2571 [astro-ph.CO]].




\bibitem{Nojiri:2019lqw}
S.~Nojiri, S.~D.~Odintsov, V.~K.~Oikonomou and T.~Paul,
Phys. Rev. D \textbf{100} (2019) no.8, 084056
doi:10.1103/PhysRevD.100.084056
[arXiv:1910.03546 [gr-qc]].




\bibitem{Elizalde:2020zcb}
E.~Elizalde, S.~D.~Odintsov, V.~K.~Oikonomou and T.~Paul,
Nucl. Phys. B \textbf{954} (2020), 114984
doi:10.1016/j.nuclphysb.2020.114984
[arXiv:2003.04264 [gr-qc]].







\bibitem{Bamba:2012ka}
K.~Bamba, J.~de Haro and S.~D.~Odintsov,
JCAP {\bf 1302} (2013) 008
doi:10.1088/1475-7516/2013/02/008
[arXiv:1211.2968 [gr-qc]].


\bibitem{Ellis:2003qz}
G.~F.~R.~Ellis, J.~Murugan and C.~G.~Tsagas,
Class. Quant. Grav. \textbf{21} (2004) no.1, 233-250
doi:10.1088/0264-9381/21/1/016
[arXiv:gr-qc/0307112 [gr-qc]].


\bibitem{Paul:2020bje}
B.~C.~Paul, S.~D.~Maharaj and A.~Beesham,
[arXiv:2008.00169 [astro-ph.CO]].


\bibitem{Li:2019laq}
S.~L.~Li, H.~L\"u, H.~Wei, P.~Wu and H.~Yu,
Phys. Rev. D \textbf{99} (2019) no.10, 104057
doi:10.1103/PhysRevD.99.104057
[arXiv:1903.03940 [gr-qc]].







\bibitem{Akrami:2018odb}
Y.~Akrami {\it et al.} [Planck Collaboration],
arXiv:1807.06211 [astro-ph.CO].


\bibitem{Nojiri:2017ncd}
S.~Nojiri, S.~D.~Odintsov and V.~K.~Oikonomou,
Phys.\ Rept.\ {\bf 692} (2017) 1 doi:10.1016/j.physrep.2017.06.001
[arXiv:1705.11098 [gr-qc]].

\bibitem{Nojiri:2010wj}
S.~Nojiri and S.~D.~Odintsov,
Phys.\ Rept.\ {\bf 505} (2011) 59
doi:10.1016/j.physrep.2011.04.001 [arXiv:1011.0544 [gr-qc]].


\bibitem{Elizalde:2018rmz}
E.~Elizalde, S.~D.~Odintsov, T.~Paul and D.~S\'aez-Chill\'on G\'omez,
Phys. Rev. D \textbf{99} (2019) no.6, 063506
doi:10.1103/PhysRevD.99.063506
[arXiv:1811.02960 [gr-qc]].




\bibitem{Perlmutter:1996ds}
S.~Perlmutter \textit{et al.} [Supernova Cosmology Project],
Astrophys. J. \textbf{483} (1997), 565
doi:10.1086/304265
[arXiv:astro-ph/9608192 [astro-ph]].


\bibitem{Perlmutter:1998np}
S.~Perlmutter \textit{et al.} [Supernova Cosmology Project],
Astrophys. J. \textbf{517} (1999), 565-586
doi:10.1086/307221
[arXiv:astro-ph/9812133 [astro-ph]].


\bibitem{Riess:1998cb}
A.~G.~Riess \textit{et al.} [Supernova Search Team],
Astron. J. \textbf{116} (1998), 1009-1038
doi:10.1086/300499
[arXiv:astro-ph/9805201 [astro-ph]].




\bibitem{Capozziello:2011et}
S.~Capozziello and M.~De Laurentis,
Phys.\ Rept.\ {\bf 509} (2011) 167
doi:10.1016/j.physrep.2011.09.003 [arXiv:1108.6266 [gr-qc]].

\bibitem{Capozziello:2010zz}
V.~Faraoni and S.~Capozziello,
Fundam.\ Theor.\ Phys.\ {\bf 170} (2010).
doi:10.1007/978-94-007-0165-6


\bibitem{Green:1984sg}
M.~B.~Green and J.~H.~Schwarz,
Phys. Lett. B \textbf{149} (1984), 117-122
doi:10.1016/0370-2693(84)91565-X

\bibitem{Antoniadis:1992sa}
I.~Antoniadis, E.~Gava and K.~S.~Narain,
Phys. Lett. B \textbf{283} (1992), 209-212
doi:10.1016/0370-2693(92)90009-S
[arXiv:hep-th/9203071 [hep-th]].

\bibitem{Hwang:2005hb}
J.~c.~Hwang and H.~Noh,
Phys. Rev. D \textbf{71} (2005), 063536
doi:10.1103/PhysRevD.71.063536
[arXiv:gr-qc/0412126 [gr-qc]].


\bibitem{Choi:1999zy}
K.~Choi, J.~c.~Hwang and K.~W.~Hwang,
Phys. Rev. D \textbf{61} (2000), 084026
doi:10.1103/PhysRevD.61.084026
[arXiv:hep-ph/9907244 [hep-ph]].


\bibitem{Satoh:2007gn}
M.~Satoh, S.~Kanno and J.~Soda,
Phys. Rev. D \textbf{77} (2008), 023526
doi:10.1103/PhysRevD.77.023526
[arXiv:0706.3585 [astro-ph]].


\bibitem{Satoh:2008ck}
M.~Satoh and J.~Soda,
JCAP \textbf{09} (2008), 019
doi:10.1088/1475-7516/2008/09/019
[arXiv:0806.4594 [astro-ph]].


\bibitem{Haghani:2017yjk}
Z.~Haghani, T.~Harko and S.~Shahidi,
Eur. Phys. J. C \textbf{77} (2017) no.8, 514
doi:10.1140/epjc/s10052-017-5078-0
[arXiv:1704.06539 [gr-qc]].


\bibitem{Nishizawa:2018srh}
A.~Nishizawa and T.~Kobayashi,
Phys. Rev. D \textbf{98} (2018) no.12, 124018
doi:10.1103/PhysRevD.98.124018
[arXiv:1809.00815 [gr-qc]].


\bibitem{Jackiw:2003pm}
R.~Jackiw and S.~Y.~Pi,
Phys. Rev. D \textbf{68} (2003), 104012
doi:10.1103/PhysRevD.68.104012
[arXiv:gr-qc/0308071 [gr-qc]].


\bibitem{Lue:1998mq}
A.~Lue, L.~M.~Wang and M.~Kamionkowski,
Phys. Rev. Lett. \textbf{83} (1999), 1506-1509
doi:10.1103/PhysRevLett.83.1506
[arXiv:astro-ph/9812088 [astro-ph]].

\bibitem{Inomata:2018rin}
K.~Inomata and M.~Kamionkowski,
Phys. Rev. Lett. \textbf{123} (2019) no.3, 031305
doi:10.1103/PhysRevLett.123.031305
[arXiv:1811.04959 [astro-ph.CO]].


\bibitem{Kamionkowski:1997av}
M.~Kamionkowski and A.~Kosowsky,
Phys. Rev. D \textbf{57} (1998), 685-691
doi:10.1103/PhysRevD.57.685
[arXiv:astro-ph/9705219 [astro-ph]].


\bibitem{Wagle:2018tyk}
P.~Wagle, N.~Yunes, D.~Garfinkle and L.~Bieri,
Class. Quant. Grav. \textbf{36} (2019) no.11, 115004
doi:10.1088/1361-6382/ab0eed
[arXiv:1812.05646 [gr-qc]].


\bibitem{Odintsov:2019mlf}
S.~D.~Odintsov and V.~K.~Oikonomou,
Phys. Rev. D \textbf{99} (2019) no.6, 064049
doi:10.1103/PhysRevD.99.064049
[arXiv:1901.05363 [gr-qc]].

\bibitem{Odintsov:2016tar}
S.~D.~Odintsov and V.~K.~Oikonomou,
Phys. Rev. D \textbf{94} (2016) no.6, 064022
doi:10.1103/PhysRevD.94.064022
[arXiv:1606.03689 [gr-qc]].


\bibitem{Nojiri:2019fft}
S.~Nojiri, S.~D.~Odintsov and V.~K.~Oikonomou,
Phys. Dark Univ. \textbf{29} (2020), 100602
doi:10.1016/j.dark.2020.100602
[arXiv:1912.13128 [gr-qc]].


\bibitem{Nojiri:2020wmh}
S.~Nojiri, S.~D.~Odintsov, V.~K.~Oikonomou and T.~Paul,
Phys. Rev. D \textbf{102} (2020) no.2, 023540
doi:10.1103/PhysRevD.102.023540
[arXiv:2007.06829 [gr-qc]].

\bibitem{Bajardi:2021hya}
F.~Bajardi, D.~Vernieri and S.~Capozziello,
[arXiv:2106.07396 [gr-qc]].


\bibitem{Lehners:2008qe}
J.~L.~Lehners and P.~J.~Steinhardt,
Phys. Rev. D \textbf{79} (2009), 063503
doi:10.1103/PhysRevD.79.063503
[arXiv:0812.3388 [hep-th]].


\bibitem{Nojiri:2005sx}
S.~Nojiri, S.~D.~Odintsov and S.~Tsujikawa,
Phys. Rev. D \textbf{71} (2005), 063004
doi:10.1103/PhysRevD.71.063004
[arXiv:hep-th/0501025 [hep-th]].

\bibitem{Aghanim:2018eyx}
N.~Aghanim \textit{et al.} [Planck],
Astron. Astrophys. \textbf{641} (2020), A6
doi:10.1051/0004-6361/201833910
[arXiv:1807.06209 [astro-ph.CO]].

\bibitem{Elizalde:2010ts}
E.~Elizalde, S.~Nojiri, S.~D.~Odintsov, L.~Sebastiani and S.~Zerbini,
Phys. Rev. D \textbf{83} (2011), 086006
doi:10.1103/PhysRevD.83.086006
[arXiv:1012.2280 [hep-th]].


\bibitem{Nojiri:2021iko}
S.~Nojiri, S.~D.~Odintsov and T.~Paul,
Symmetry \textbf{13} (2021) no.6, 928
doi:10.3390/sym13060928
[arXiv:2105.08438 [gr-qc]].


\bibitem{new1}
V.A. Belinskii, I.M. Khalatnikov and E.M. Lifshitz ; Advances in Physics 19, 525 (1970).

\bibitem{Erickson:2003zm}
  J.~K.~Erickson, D.~H.~Wesley, P.~J.~Steinhardt and N.~Turok,
  Phys.\ Rev.\ D {\bf 69} (2004) 063514
  doi:10.1103/PhysRevD.69.063514
  [hep-th/0312009].
  
  
  
\bibitem{Garfinkle:2008ei}
  D.~Garfinkle, W.~C.~Lim, F.~Pretorius and P.~J.~Steinhardt,
  Phys.\ Rev.\ D {\bf 78} (2008) 083537
  doi:10.1103/PhysRevD.78.083537
  [arXiv:0808.0542 [hep-th]].
  
  

\bibitem{Odintsov:2020nwm}
S.~D.~Odintsov and V.~K.~Oikonomou,
Phys. Rev. D \textbf{101} (2020) no.4, 044009
doi:10.1103/PhysRevD.101.044009
[arXiv:2001.06830 [gr-qc]].

\bibitem{Nojiri:2020pqr}
  S.~Nojiri, S.~D.~Odintsov, V.~K.~Oikonomou and A.~A.~Popov,
  Phys.\ Dark Univ.\  {\bf 28} (2020) 100514
  doi:10.1016/j.dark.2020.100514
  [arXiv:2002.10402 [gr-qc]].


\bibitem{Noh:2001ia}
H.~Noh and J.~c.~Hwang,
Phys.\ Lett.\ B {\bf 515} (2001) 231
doi:10.1016/S0370-2693(01)00875-9
[astro-ph/0107069].


\bibitem{Hwang:2002fp}
J.~c.~Hwang and H.~Noh,
Phys.\ Rev.\ D {\bf 66} (2002) 084009
doi:10.1103/PhysRevD.66.084009
[hep-th/0206100].



  
\bibitem{Cai:2009fn}
Y.~F.~Cai, W.~Xue, R.~Brandenberger and X.~Zhang,
JCAP \textbf{05} (2009), 011
doi:10.1088/1475-7516/2009/05/011
[arXiv:0903.0631 [astro-ph.CO]].
  
  
  
\bibitem{Agullo:2020cvg}
I.~Agullo, D.~Kranas and V.~Sreenath,
Class. Quant. Grav. \textbf{38} (2021) no.6, 065010
doi:10.1088/1361-6382/abc521
[arXiv:2006.09605 [astro-ph.CO]].

\bibitem{LIGOScientific:2016jlg}
B.~P.~Abbott \textit{et al.} [LIGO Scientific and Virgo],
Phys. Rev. Lett. \textbf{118} (2017) no.12, 121101
[erratum: Phys. Rev. Lett. \textbf{119} (2017) no.2, 029901]
doi:10.1103/PhysRevLett.118.121101
[arXiv:1612.02029 [gr-qc]].

\bibitem{Janssen:2014dka}
G.~Janssen, G.~Hobbs, M.~McLaughlin, C.~Bassa, A.~T.~Deller, M.~Kramer, K.~Lee, C.~Mingarelli, P.~Rosado and S.~Sanidas, \textit{et al.}
PoS \textbf{AASKA14} (2015), 037
doi:10.22323/1.215.0037
[arXiv:1501.00127 [astro-ph.IM]].

\bibitem{Amaro-Seoane:2012aqc}
P.~Amaro-Seoane, S.~Aoudia, S.~Babak, P.~Binetruy, E.~Berti, A.~Bohe, C.~Caprini, M.~Colpi, N.~J.~Cornish and K.~Danzmann, \textit{et al.}
GW Notes \textbf{6} (2013), 4-110
[arXiv:1201.3621 [astro-ph.CO]].










\end{thebibliography}
\end{document}